\documentclass[11pt]{article}
\usepackage{amsmath}
\usepackage[dvips]{graphicx}
\begin{document}
\title{A General Relativistic Model for the Gravitational Field of Active 
Galactic Nuclei Surrounded by a Disk}
\author{D. Vogt\thanks{e-mail: danielvt@ifi.unicamp.br}\\
Instituto de F\'{\i}sica Gleb Wataghin, Universidade Estadual de Campinas\\
13083-970 Campinas, S.\ P., Brazil
\and
P. S. Letelier\thanks{e-mail: letelier@ime.unicamp.br}\\
Departamento de Matem\'{a}tica Aplicada-IMECC, Universidade Estadual\\
de Campinas 13083-970 Campinas, S.\ P., Brazil}
\maketitle
\begin{abstract}
An exact but simple general relativistic model for the gravitational 
field of active galactic nuclei is 
constructed, based on the superposition in Weyl coordinates of a black hole, a Chazy-Curzon 
disk and two rods, which represent matter jets. The influence of the rods 
on the matter properties of the disk and on its stability is examined. We find  
that in general they contribute to destabilize the disk. Also the oscillation 
frequencies for perturbed circular geodesics on the disk are computed, and some 
geodesic orbits for the superposed metric are numerically calculated.   


\end{abstract}

\section{Introduction}

There is a strong observational evidence that active galactic nuclei (AGN), X-ray
transients and gamma-ray bursts (GRBs) are associated with accretion onto 
black holes, and that these sources are able to form collimated, ultrarelativistic
flows (relativistic jets).

The exact mechanisms to explain the production of jets are still uncertain, but they
probably involve the interaction between a spinning black hole, the accretion disk 
and electromagnetic fields in strong gravitational fields (see, for example, \cite{Kundt}-\cite{Blandford} 
and references therein). 

Thus, a reasonably accurate general relativistic model of an AGN would require an exact solution of
Einstein-Maxwell field equations that describes a superposition of a Kerr
black hole with a stationary disk and electromagnetic fields. Not even an 
exact solution of a stationary black hole-disk system has been found yet. Solutions for static thin 
disks without radial pressure were first studied by
Bonnor and Sackfield \cite{Bonnor1}, and Morgan and Morgan \cite{Morgan1}, and with 
radial pressure by Morgan and Morgan \cite{Morgan2}. Several classes of exact solutions 
of the Einstein field equations corresponding to static thin disks with or without radial 
pressure have been obtained by different authors \cite{Lynden-Bell}-\cite{Garcia}. Thin
rotating disks were considered in \cite{Bicak3, Gonzalez1}. Perfect
fluid disks with halos \cite{Vogt1} and charged perfect fluid disks \cite{Vogt2} were 
also studied, whereas thick relativistic disks were reported in
\cite{Gonzalez3}. Several solutions  of  the Einstein equations coupled to
matter that represent disks has also been studied  by the
 Jena group  \cite{Klein1}-\cite{Klein6}.

The static
superposition of a disk and a black hole was first considered by Lemos and 
Letelier \cite{Lemos2}. Zellerin and Semer\'{a}k \cite{Zellerin} found a stationary metric that 
reduces to the superposition of a disk and a
black hole in the static limit and thus may represent a stationary disk-black hole system. The 
analysis of their solution is complicated by the fact that the metric functions cannot be 
analytically computed. For a survey on self gravitating relativistic thin disks, see for instance \cite{Karas}.

The aim of this paper is to consider the gravitational field of an 
AGN through a simple model: the \emph{static} superposition of a black hole with a Chazy-Curzon 
disk and two rods placed on the symmetry axis, which will represent jets. Our 
principal interest here is to see how the presence of the rods affect the 
matter properties and stability of the disk.

The article is divided as follows. In Sec.\ \ref{sec_disk} we review 
the ``displace, cut and reflect'' method used to construct thin disks 
from a known solution of Einstein field equations in Weyl coordinates. Sec.\ \ref{sec_sup} 
summarizes the formalism to superpose thin disks and other Weyl solutions. Sec.\ \ref{sec_sol} 
discusses Schwarzschild solution and the metric of a finite rod in Weyl 
coordinates. In Sec.\ \ref{sec_model} the results of Sec.\ \ref{sec_sup}  and \ref{sec_sol}  are then applied to 
construct the superposition of disk, black hole and rods and the resulting 
energy-momentum tensor. In Sec.\ \ref{sec_osc} the disk stability is studied through 
small horizontal and vertical oscillations about equatorial circular 
geodesics. In Sec.\ \ref{sec_orbits} some geodesic orbits for the superposed metric are numerically calculated. Finally, 
Sec.\ \ref{sec_discuss} is devoted to discussion of the results. We take units such
that $c=G=1$. 
\section{Thin disk solutions in Weyl coordinates} \label{sec_disk}

In absense of matter, the general metric for a static axially symmetric spacetime in
Weyl's canonical coordinates $(t,r,z,\varphi)$ is given by 
\begin{equation} \label{eq_weyl_metric}
\mathrm{d}s^2=-e^{\phi}\mathrm{d}t^2+e^{\nu-\phi}(\mathrm{d}r^2+\mathrm{d}z^2)+
r^2e^{-\phi}\mathrm{d}\varphi^2 \mbox{,}
\end{equation}
where $\phi$ and $\nu$ are functions of $r$ and $z$ only. Einstein vacuum field 
equations for the metric Eq.\ (\ref{eq_weyl_metric}) yield \cite{Weyl1,Weyl2}
\begin{subequations}
\begin{align}
&\phi_{,rr}+\frac{\phi_{,r}}{r}+\phi_{,zz}=0 \label{eq_eins1} \mbox{,} \\
&\nu[\phi]=\frac{1}{2}\int r \left[ (\phi_{,r}^2-\phi_{,z}^2)\mathrm{d}r+
2\phi_{,r}\phi_{,z}\mathrm{d}z \right] \label{eq_eins2} \mbox{.}
\end{align}
\end{subequations}
Given a solution of Eq.\ (\ref{eq_eins1})-(\ref{eq_eins2}), one can construct 
a thin disk by using the well known ``displace, cut and reflect'' method, due to Kuzmin \cite{Kuzmin}.
First, a surface $(z=0)$ is chosen so that it divides the usual space in two parts: 
one with no singularities or sources, and the other with them. Then the part 
of the space with singularities or sources is disregarded. At last, the surface 
is used to make an inversion of the nonsingular part of the space. The result 
will be a space with a singularity that is a delta function with support on 
$z=0$. The method is mathematically equivalent to make a transformation 
$z \rightarrow |z|+a$, where $a$ is a constant.

The application of the formalism of distributions in curved spacetimes to 
the Weyl metric Eq.\ (\ref{eq_weyl_metric}) is exposed in \cite{Lemos3}. One finds that
the components of the distributional energy-momentum tensor $[T^a{}_b]$ on the 
disk are
\begin{align}
-T^t{}_t&=e^{\phi-\nu}(2-r\phi_{,r})\phi_{,z}\delta(z)\mbox{,} 
\label{eq_T_tt}\\
T^{\varphi}{}_{\varphi}&=e^{\phi-\nu}r\phi_{,r}\phi_{,z}\delta(z) \mbox{,}
\label{eq_T_phi}\\
T^r{}_r&=T^z{}_z=0 \mbox{,}
\end{align}
where $\delta(z)$ is the Dirac distribution with support on the disk and is 
understood that $\phi_{,z}=\lim_{z \rightarrow 0^+} \phi_{,z}$.
The ``true'' energy density and azimuthal pressure are, respectively, 
\begin{align}
\sigma &=e^{(\nu-\phi)/2}(-T^t{}_t)\mbox{,} \label{eq_sigma} \\
p &=e^{(\nu-\phi)/2}T^{\varphi}{}_{\varphi} \mbox{.} \label{eq_p}
\end{align}

To explain the disk stability in absence of radial pressure, one may assume the 
counterrotating hypothesis, where the particles on the disk move in such a way
that there are as many particles moving in clockwise as in counterckockwise
direction. The velocity $V$ of counterrotation of the particles in the disk is given
by \cite{Lemos1, Lemos5}
\begin{equation} \label{eq_vel}
V^2=\frac{p}{\sigma}
\end{equation}
If $V^2<1$, the particles travel at sublumial velocities. The specific angular 
momentum $h$ of particles on the disk moving in circular orbits along geodesics 
reads
\begin{equation} \label{eq_ang_mom}
h=r^{3/2}e^{-\phi/2}\sqrt{\frac{\phi_{,r}}{2(1-r\phi_{,r})}} \mbox{.}
\end{equation}
The stability of circular orbits on the disk plane can be determined with an 
extension of Rayleigh criteria of stability of a fluid at rest in a gravitational 
field: $h\frac{\mathrm{d}h}{\mathrm{d}r}>0$ \cite{Rayleigh}. Using Eq.\ (\ref{eq_ang_mom}) this 
is equivalent to
\begin{equation} \label{eq_stab}
\phi_{,r}(-3r\phi_{,r}+3+r^2\phi_{,r}^2)+r\phi_{,rr}>0 \mbox{.}
\end{equation}

\section{Superposition of thin disks and other Weyl solutions} \label{sec_sup}

An important property of the Weyl metric Eq.\ (\ref{eq_weyl_metric}) is that 
the field equation (\ref{eq_eins1}) for the potential $\phi$ is the Laplace 
equation in cylindrical coordinates. Since Laplace's equation is linear, if 
$\phi_1$ and $\phi_2$ are solutions, then the superposition $\phi=\phi_1+\phi_2$ 
is also a solution. The other metric function Eq.\ (\ref{eq_eins2}) is nonlinear, and so 
cannot be superposed. But one can show that the relation
\begin{equation} \label{eq_nu_rel1}
\nu[\phi_1+\phi_2]=\nu[\phi_1]+\nu[\phi_2]+2\nu[\phi_1,\phi_2] \mbox{,}
\end{equation}
where
\begin{equation}
\nu[\phi_1,\phi_2]=\frac{1}{2}\int r[(\phi_{1,r}\phi_{2,r}-\phi_{1,z}
\phi_{2,z})\mathrm{d}r+(\phi_{1,r}\phi_{2,z}+\phi_{1,z}\phi_{2,r})\mathrm{d}z]
\mbox{,}
\end{equation}
holds. Other useful relations are given in \cite{Letelier1}.

The energy-momentum tensor of the combined system disk and black hole has
been computed by Lemos and Letelier \cite{Lemos3}. Let $\phi_D$ and $\phi_{BH}$ be
the metric potentials of the disk and of the black hole, respectively. Then the 
components $[T^a{}_b]$ of the superposition are
\begin{align}
-T^t{}_t &=e^{\phi_{D}+\phi_{BH}-\nu}\left[ 2-r(\phi_{D}+\phi_{BH})_{,r}
\right] \phi_{D,z} \delta(z) \mbox{,} \label{eq_Ttt_sup} \\
T^{\varphi}{}_{\varphi} &=e^{\phi_{D}+\phi_{BH}-\nu}r(\phi_{D}+\phi_{BH})_{,r}
\phi_{D,z} \delta(z) \mbox{,} \label{eq_Tphi_sup} \\
T^r{}_r &=T^z{}_z=0 \mbox{,}
\end{align}
where $\nu=\nu[\phi_{D}+\nu_{BH}]$, and again $\phi_{D,z}=\lim_{z \rightarrow 0^+} 
\phi_{D,z}$. The ``true'' energy density and azimuthal pressure read
\begin{align}
\sigma &=e^{(\nu-\phi_{D}-\phi_{BH})/2}(-T^t{}_t)\mbox{,} \label{eq_sigma_sup} \\
p &=e^{(\nu-\phi_{D}-\phi_{BH})/2}T^{\varphi}{}_{\varphi} \mbox{.} \label{eq_p_sup}
\end{align}
Eq.\ (\ref{eq_Ttt_sup})-(\ref{eq_Tphi_sup}) show that the potential of the black 
hole interacts with the disk and changes its matter properties. Although Eq.\ 
(\ref{eq_Ttt_sup})-(\ref{eq_p_sup}) have been derived for superposition of disk and black 
hole, they are also valid when the potential function $\phi_{BH}$ is a sum of other 
Weyl solutions, like the superposition of a black hole and rods.

\section{Black holes and rods in Weyl coordinates} \label{sec_sol}

The Schwarzschild black hole metric function $\phi_{BH}$ in Weyl coordinates is given by
\begin{equation}\label{eq_sch1}
\phi_{BH} =\ln \left( \frac{r_1+r_2-2M}{r_1+r_2+2M} \right) \mbox{,}
\end{equation}
where $r_1^2=(M-z)^2+r^2$ and $r_2^2=(M+z)^2+r^2$. 
The function $\phi(r,z)$ can be related to the Newtonian potential $U$ by 
\begin{equation}
\phi=2U \mbox{.}
\end{equation}
Thus, the metric potential $\phi_R$ of a finite rod of linear mass density $\lambda$ lying on
the $z$ axis and located along $[c_1,c_2]$ is
\begin{equation} \label{eq_bar1}
\phi_R=-2\lambda \ln \left[ \frac{c_2-z+\sqrt{r^2+(c_2-z)^2}}
{c_1-z+\sqrt{r^2+(c_1-z)^2}} \right] \mbox{.}
\end{equation}
The calculation of the other metric function $\nu$ for Eq.\ (\ref{eq_sch1}), (\ref{eq_bar1}) 
and later for the superposed metric, is considerably simplified when one defines the following 
$\mu$ function
\begin{equation}
\mu_k=\alpha_k-z+\sqrt{r^2+(\alpha_k-z)^2} \mbox{,}
\end{equation}
where $\alpha_k$ is an arbitrary constant. This function is a natural consequence 
of the formalism of the inverse scattering method \cite{Belinsky1,Belinsky2}. Eq.\ (\ref{eq_sch1})
and (\ref{eq_bar1}) can be rewritten as
\begin{align}
\phi_{BH} &=\ln \left( \frac{\mu_1}{\mu_2} \right) \mbox{,} \\
\phi_R &= -2\lambda \ln \left( \frac{\mu_3}{\mu_4} \right) \mbox{,}
\end{align}
where we defined
\begin{align*}
\mu_1 &=-M-z+\sqrt{r^2+(M+z)^2} \mbox{,}
& \mu_2 =M-z+\sqrt{r^2+(M-z)^2} \mbox{,} \\
\mu_3 &=c_2-z+\sqrt{r^2+(c_2-z)^2} \mbox{,}
& \mu_4 =c_1-z+\sqrt{r^2+(c_1-z)^2} \mbox{.}
\end{align*}
On using Eq.\  (\ref{eq_nu_rel1})
\begin{equation} \label{eq_nu_rel2}
\nu[\ln \mu_i-\ln \mu_j]=\nu[\ln \mu_i]+\nu[\ln \mu_j]-2\nu[\ln \mu_i,\ln \mu_j] \mbox{;}
\end{equation}
the result
\begin{equation}
\nu[\ln \mu_i,\ln \mu_j]=\ln (\mu_i-\mu_j) \mbox{,}
\end{equation}
which also follows from the inverse scattering method; and the identity
\begin{equation} \label{eq_ident}
(r^2+\mu_i \mu_j)(\mu_i-\mu_j)=2(\alpha_i-\alpha_j)\mu_i\mu_j \mbox{,}
\end{equation}
one obtains following expressions for the metric function $\nu$
\begin{align}
\nu_{BH} &=\ln \left[ \frac{(r^2+\mu_1 \mu_2)^2}{(r^2+\mu_1^2)(r^2+\mu_2^2)} \right] \mbox{,} \label{eq_nu_bh}\\
\nu_{R} &= 4\lambda^2 \ln \left[ \frac{(r^2+\mu_3 \mu_4)^2}{(r^2+\mu_3^2)(r^2+\mu_4^2)} 
\right] \mbox{.} \label{eq_nu_rod}
\end{align}
\section{Superposition of disk, black hole and rods} \label{sec_model}

\begin{figure}
\centering
\includegraphics[scale=0.8]{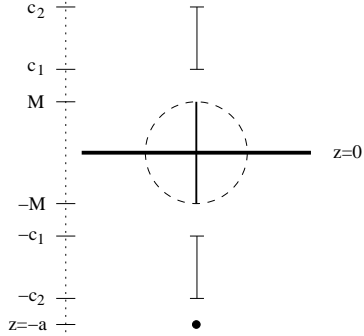}
\caption{Superposition of a black hole with mass $M$, two rods and a Chazy-Curzon
disk on the plane $z=0$.} \label{fig_1}
\end{figure}

We now consider the superposition illustrated in Fig.\ \ref{fig_1}: a black hole with mass 
$M$ whose center is on $z=0$, two rods with equal mass density $\lambda$, each one with mass $\mathcal{M}$
located along $[-c_2,-c_1]$ and $[c_1,c_2]$ on the $z$ axis, and a disk on the plane $z=0$ constructed with
the ``displace, cut and reflect method'' from the Chazy-Curzon solution with mass $m$, whose 
singularity lies on $z=-a$:
\begin{equation}\label{eq_curzon}
\phi_D=-\frac{2m}{\sqrt{r^2+(|z|+a)^2}} \mbox{.}
\end{equation}
It should be remembered that in Weyl coordinates a black hole with mass $M$ is represented by 
a rod with length $2M$, thus in Fig.\ \ref{fig_1} we put a dotted circle around the rod in the middle.
Such a configuration is not gravitationally stable: a consequence of the nonlinearity of Eq.\ (\ref{eq_eins2})
is the appearence of gravitationally inert singular structures like struts between the rods and the 
black hole that keeps them appart; also in the superposition 
of the disk with the black hole, superlumial regions $(V^2>1)$ exist because there is matter up 
to the event horizon. 

The metric function $\phi$ of the superposition can be expressed as
\begin{equation} \label{eq_phi_sup}
\phi=-2\lambda \ln \left( \frac{\mu_3}{\mu_4} \right)+\ln \left( \frac{\mu_1}{\mu_2} \right)
-2\lambda \ln \left( \frac{\mu_5}{\mu_6} \right)+\phi_D \mbox{,}
\end{equation}
with $\mu_5=-c_1-z+\sqrt{r^2+(c_1+z)^2}$ and $\mu_6=-c_2-z+\sqrt{r^2+(c_2+z)^2}$.

Now we consider the case when both rods just touch the horizon of the black hole, that is, 
when $c_1=M$. Then $\mu_4=\mu_2$ and $\mu_5=\mu_1$. From Eq.\ (\ref{eq_Ttt_sup})-(\ref{eq_Tphi_sup}) 
and Eq.\ (\ref{eq_vel}), we get following conditions:
\begin{align}
\sigma & >0 \rightarrow \sqrt{\tilde{r}^2+\tilde{c}_2^2}\left[ (\tilde{r}^2+\tilde{a}^2)^{3/2}
(\sqrt{1+\tilde{r}^2}-1)-\alpha \tilde{r}^2\sqrt{1+\tilde{r}^2} \right] \notag \\
&+2\lambda (\tilde{r}^2+\tilde{a}^2)^{3/2} \left( \sqrt{\tilde{r}^2+\tilde{c}_2^2} -\tilde{c}_2\sqrt{1+\tilde{r}^2}
\right) >0 \mbox{,} \label{eq_cond_sigma} \\
p & >0 \rightarrow \sqrt{\tilde{r}^2+\tilde{c}_2^2} \left[\alpha \tilde{r}^2\sqrt{1+\tilde{r}^2} +
(\tilde{r}^2+\tilde{a}^2)^{3/2} \right] \notag \\
&-2\lambda (\tilde{r}^2+\tilde{a}^2)^{3/2} \left( 
\sqrt{\tilde{r}^2+\tilde{c}_2^2} -\tilde{c}_2\sqrt{1+\tilde{r}^2} \right) >0 \mbox{,} \label{eq_cond_p} \\
V^2 &<1 \rightarrow \sqrt{\tilde{r}^2+\tilde{c}_2^2} \left[ 2\alpha \tilde{r}^2\sqrt{1+\tilde{r}^2} +
(\tilde{r}^2+\tilde{a}^2)^{3/2}(2-\sqrt{1+\tilde{r}^2}) \right] \notag \\
&-4\lambda (\tilde{r}^2+\tilde{a}^2)^{3/2}
\left(\sqrt{\tilde{r}^2+\tilde{c}_2^2}-\tilde{c}_2\sqrt{1+\tilde{r}^2} \right) <0 \mbox{,}
\label{eq_cond_V}
\end{align}
where $\tilde{r}=r/M$, $\tilde{a}=a/M$, $\tilde{c}_2=c_2/M$, $\alpha=m/M$, 
$\beta=\mathcal{M}/M$ and $\lambda=\mathcal{M}/(c_2-M)=\beta /(\tilde{c}_2-1)$. The 
conditions imposed are that of weak energy $(\sigma >0)$, azimuthal pressure $(p>0)$ 
and sublumial velocity $(V^2<1)$ of counterrotation of particles on the disk.
For $\tilde{r} \rightarrow \infty$, all three conditions are satisfied. In the regions where $V^2<1$, 
the weak energy condition is always satisfied, as can be seen by inequalities 
(\ref{eq_cond_sigma}) and (\ref{eq_cond_V}).
\begin{figure}
\centering
\includegraphics[scale=0.8]{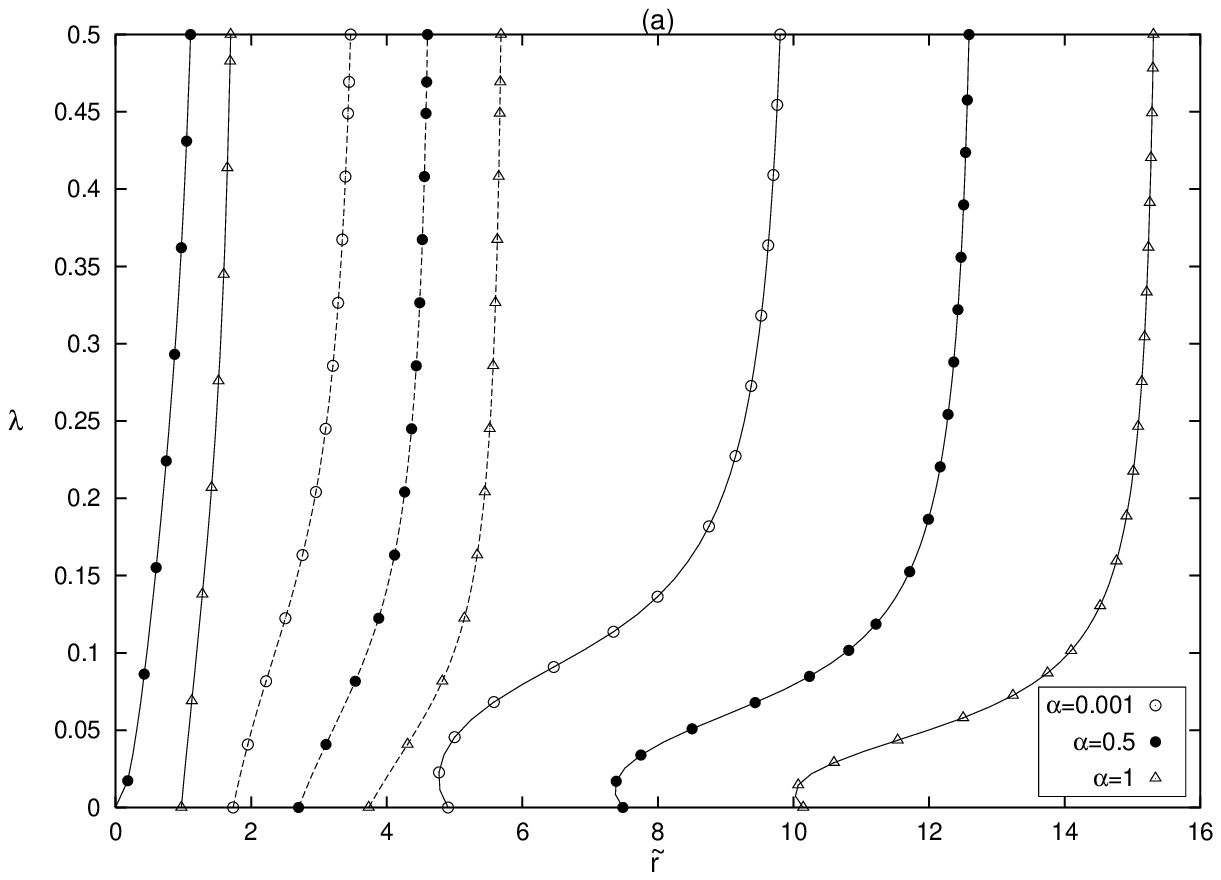}\\
\vspace{0.2cm}
\includegraphics[scale=0.8]{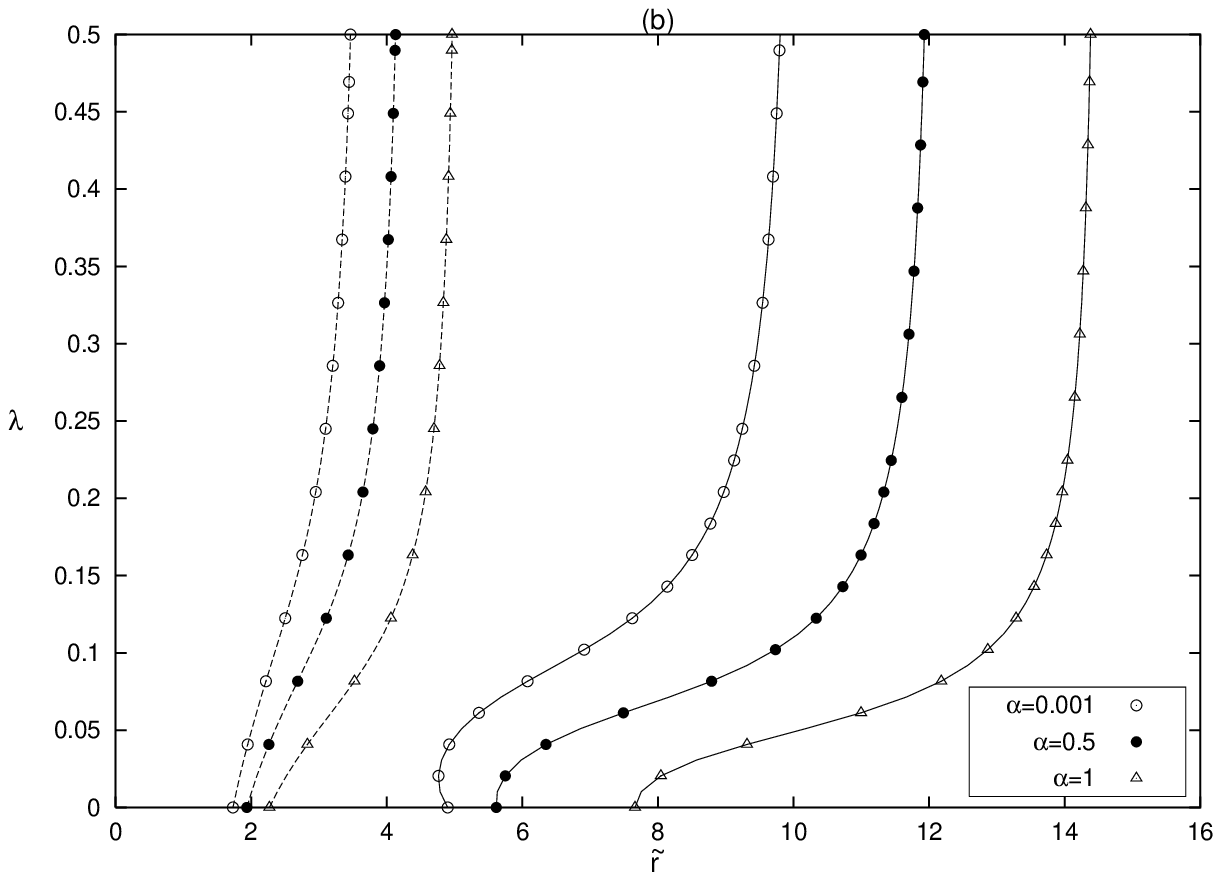}
\caption{Curves of $V^2=1$ (dotted curves) and of $h dh/dr=0$ (solid curves) for the
Chazy-Curzon disk in presence of a black hole and two rods. We keep 
the mass of each rod constant $\beta=0.5$ and vary its mass density 
$\lambda$. In (a)-(b) we take, respectively, $\tilde{a}=1$ and $\tilde{a}=3$.} \label{fig_2}
\end{figure}

Fig.\ \ref{fig_2}(a)-(c) shows curves of
$V^2=1$ (dotted curves) and of  $h dh/dr=0$ (solid curves) as
function of $\lambda$ and $\tilde{r}$ for three different values of $\alpha$. The
mass of each rod is kept constant $\beta=0.5$ and the
cut parameter takes values $\tilde{a}=1$ and $\tilde{a}=3$ in (a) and (b), respectively.
At the right of each dotted curve we have $V^2<1$ and the unstable 
regions of the disk appear between the curves of $h dh/dr=0$. We note that in general with 
increasing mass of the disk and smaller length of the rods, the disk becomes more 
unstable and the regions of superlumial velocity also increase. There is, however, 
an interval of values for the rod's mass density where the zone of stability is increased, as 
can be seen in the lower part of the curve $h dh/dr=0$ for $\alpha=0.001$ in 
Fig.\ \ref{fig_2}(a). This is probably due to the prolate quadrupole moment of the rods, 
which scale as $~\mathcal{M}l^2$, where $l$ is their length. Thus, for larger rods, the 
effect of prolate deformations may overwhelm the effect of the oblate quadrupole moment 
of the disk, and increase stability (see \cite{Letelier2} for a detailed discussion of 
the effect of quadrupolar fields on the stability of circular orbits). 

\begin{figure}
\centering
\includegraphics[scale=0.8]{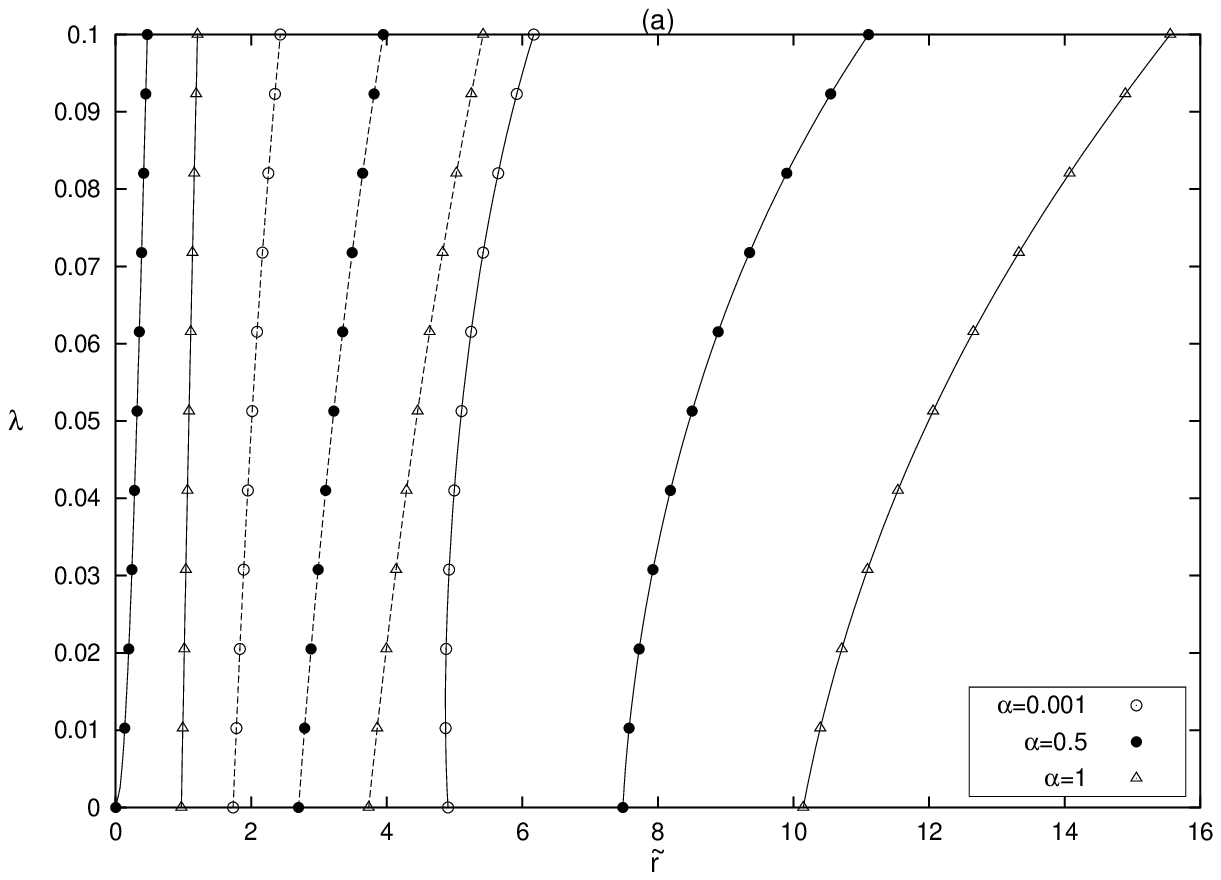}\\
\vspace{0.2cm}
\includegraphics[scale=0.8]{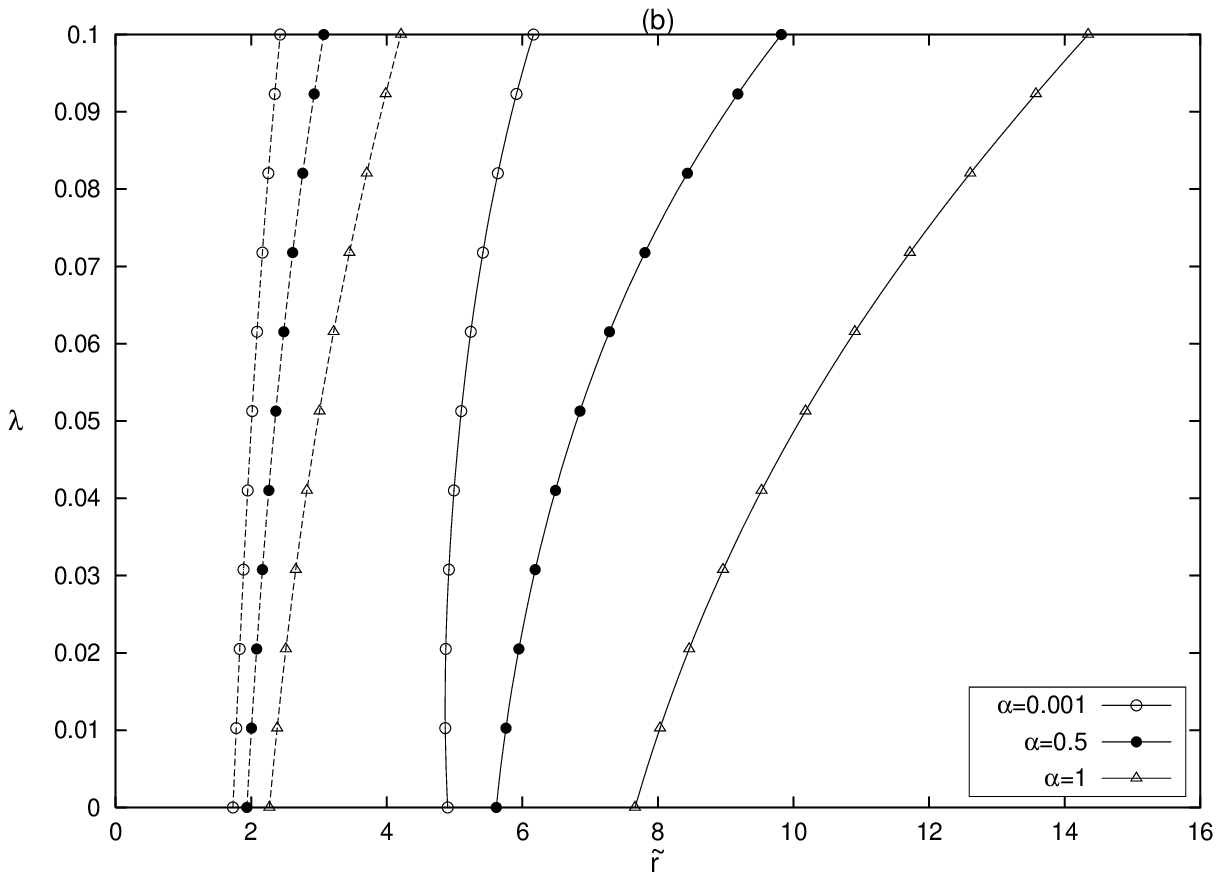}
\caption{Curves of $V^2=1$ (dotted curves) and of $h dh/dr=0$ (solid curves) for the 
Chazy-Curzon disk in presence of a black hole and two rods. We keep 
the length of each rod constant $\tilde{c}_2=11$ and vary its mass density
$\lambda$ (or equivalent, its mass). In (a)-(b) we take, respectively, $\tilde{a}=1$ and $\tilde{a}=3$.} \label{fig_3}
\end{figure}
\begin{figure}
\centering
\includegraphics[scale=0.6]{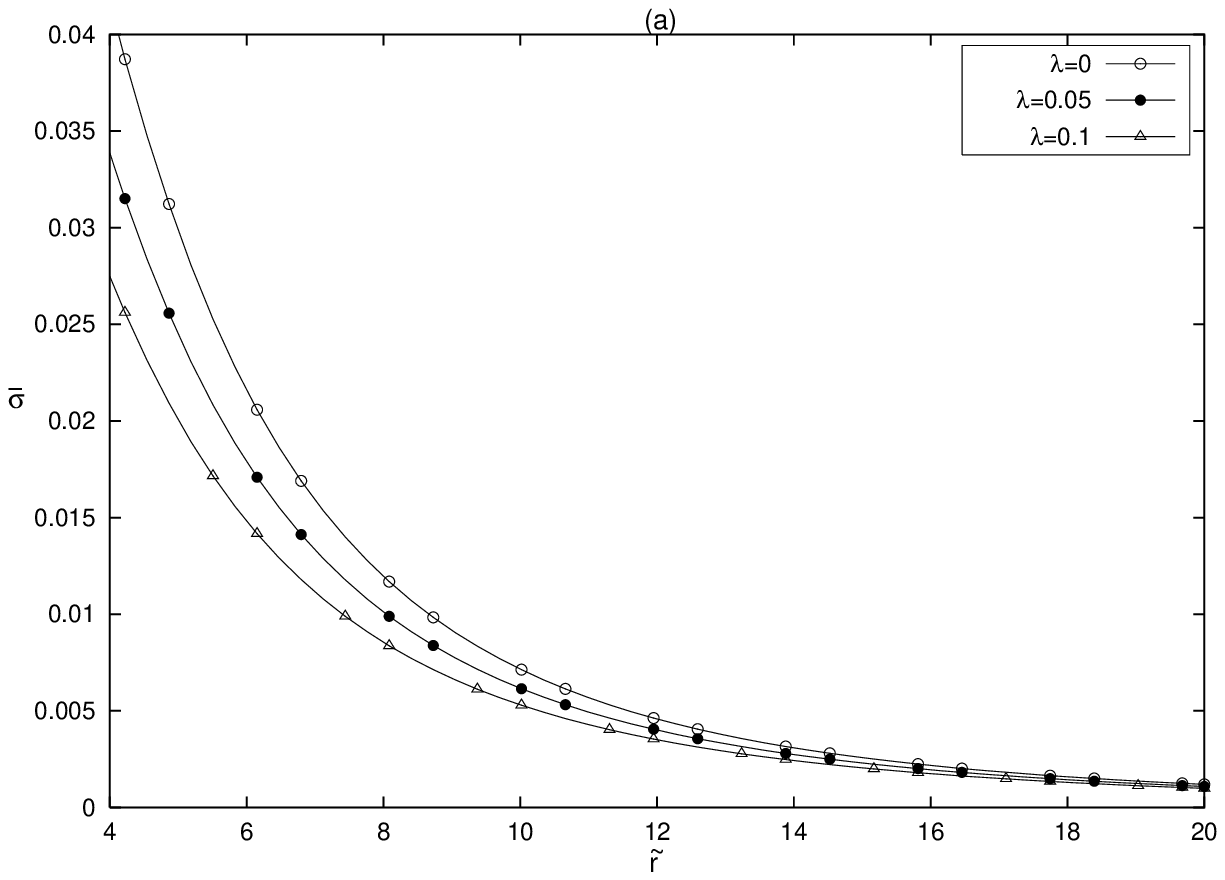}%
\includegraphics[scale=0.6]{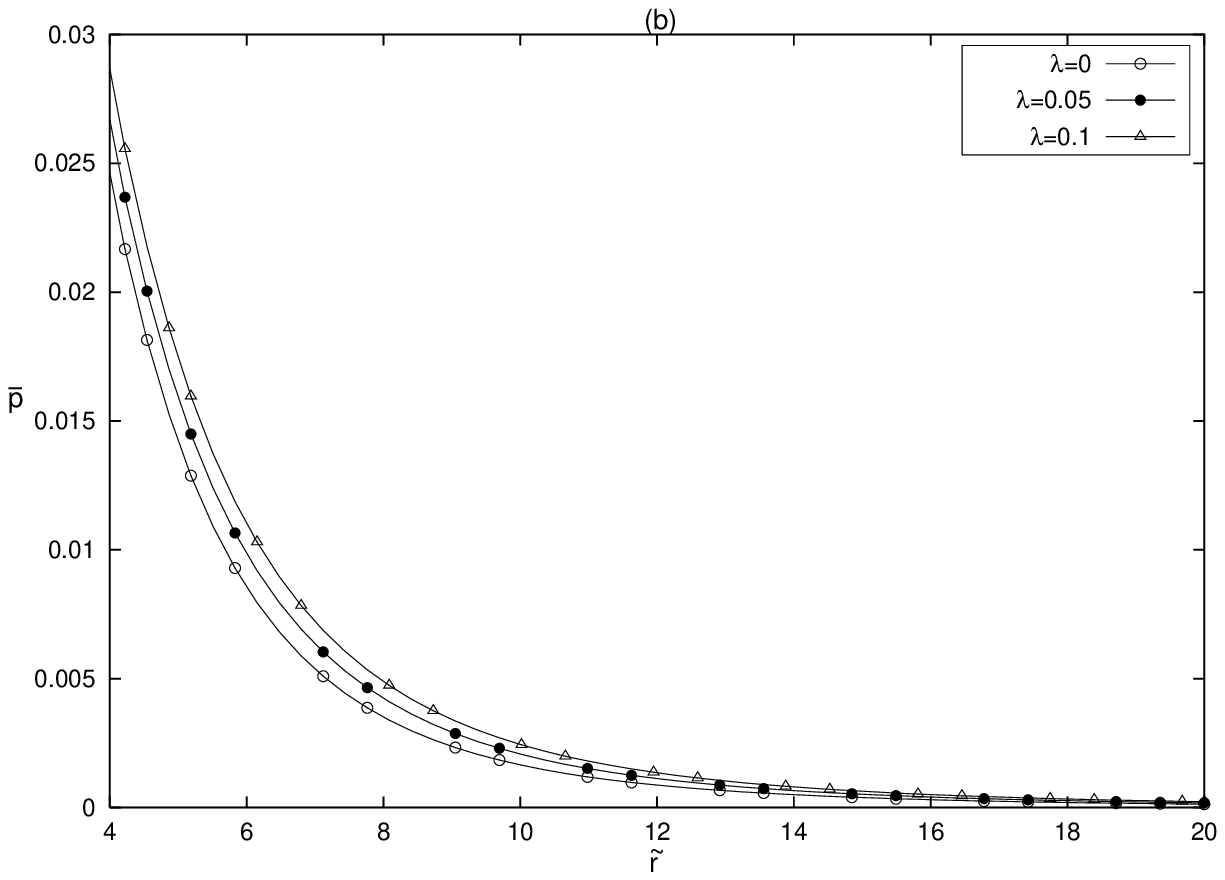} \\
\vspace{0.2cm}
\includegraphics[scale=0.6]{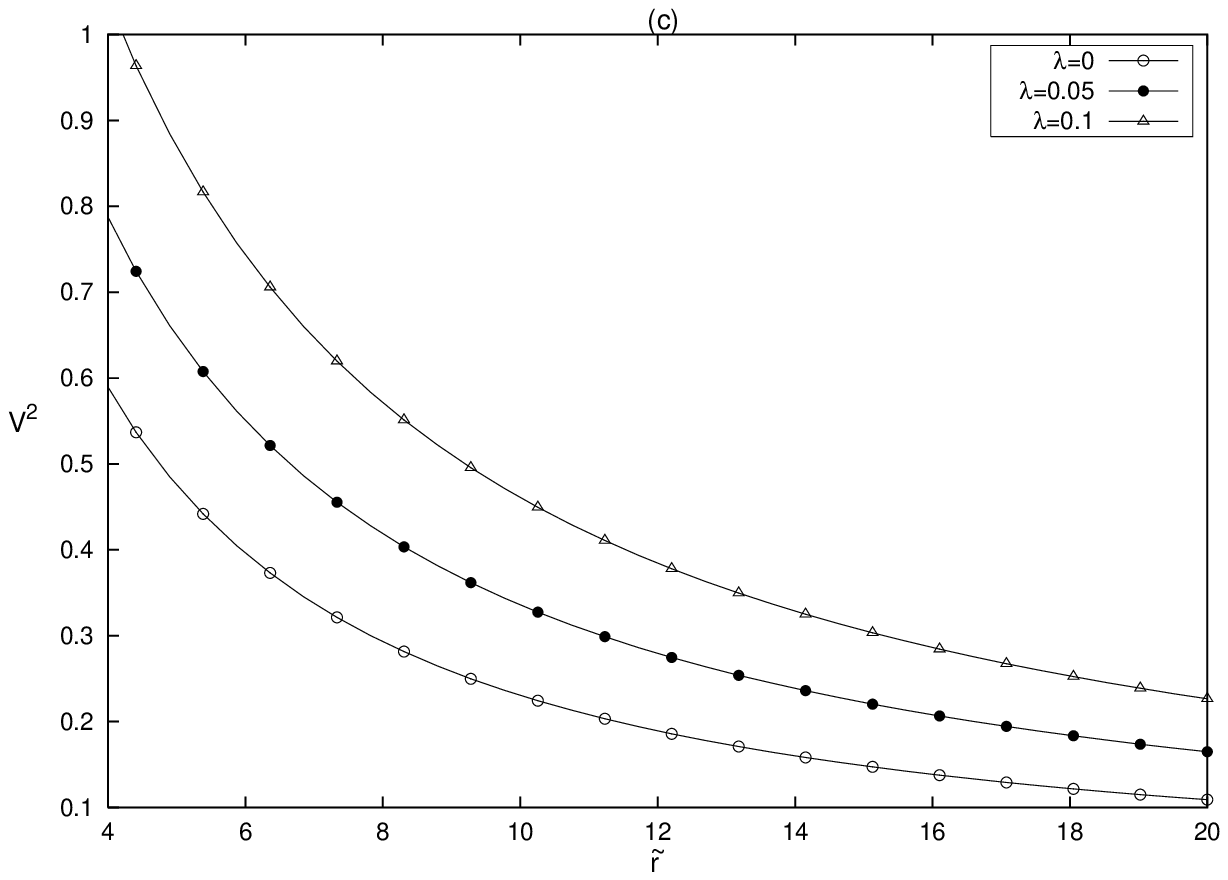}%
\includegraphics[scale=0.6]{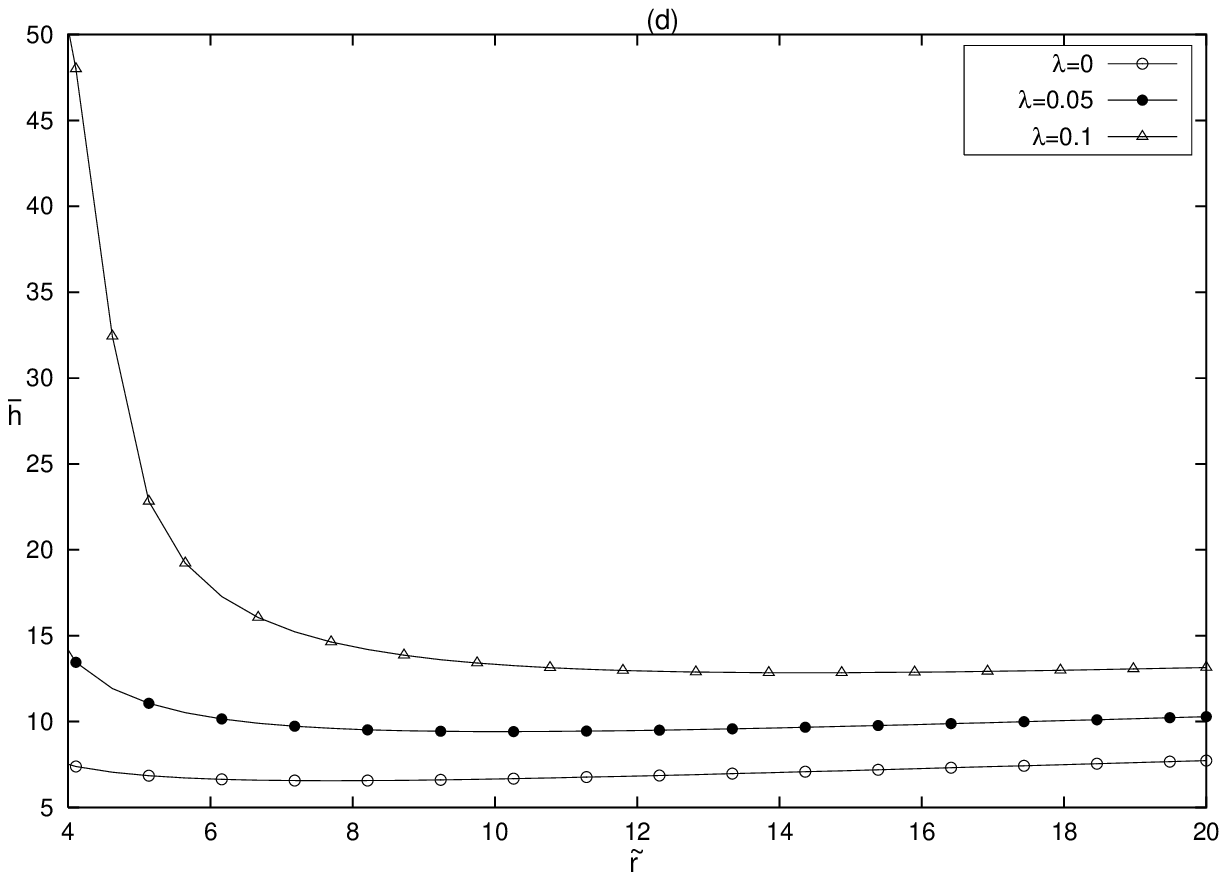}
\caption{(a) Energy density $\bar{\sigma}$, (b) azimuthal pressure $\bar{p}$,
(c) counterrotating velocity $V^2$ and (d) specific angular momentum $\bar{h}$ as 
functions of $\tilde{r}$ for $\alpha=1$, $\tilde{a}=3$, $\tilde{c}_2=11$ and three 
different values for the rod's mass density $\lambda$.} \label{fig_4}
\end{figure}

Fig.\ \ref{fig_3}(a)-(b) shows again curves of
$V^2=1$ (dotted curves) and of  $h dh/dr=0$ (solid curves) as
function of $\lambda$ and $\tilde{r}$ for three different values of $\alpha$, but 
now the length of each rod is kept constant $(\tilde{c}_2=11)$ and the
cut parameter takes values $\tilde{a}=1$ and $\tilde{a}=3$ in (a) and (b), respectively. With
increasing masses of the disk and of the rods, the zones of instability and superlumial 
velocity of the disk are enhanced. 
In Fig.\ \ref{fig_4} (a)-(d) we plot the energy density $\bar{\sigma}=M\sigma$, 
azimuthal pressure $\bar{p}=Mp$, square of the counterrotating velocity $V$ and specific angular momentum 
$\bar{h}=Mh$ as functions of $\tilde{r}$ for $\tilde{a}=3$, $\alpha=1$, 
$\tilde{c}_2=11$ (constant length) and different values of the rod's linear 
mass density $\lambda$. The curves were computed using 
Eq.\ (\ref{eq_Ttt_sup})-(\ref{eq_p_sup}), Eq.\ (\ref{eq_vel})-(\ref{eq_ang_mom})
and Eq.\ (\ref{eq_phi_sup}). The expression for the corresponding metric 
function $\nu$ is given in the Appendix. Energy density is lowered for a 
fixed radius as the rods become more massive, while pressure is slightly 
increased. Velocity of counterrotation and specific angular momentum are 
enhanced by increasing mass of the rods, as can also be deduced from Fig.\ \ref{fig_3}(b).
\section{Horizontal and vertical oscillations of the disk} \label{sec_osc}

It is interesting to study the disk stability through the computation of horizontal (epicyclic)
and vertical oscillation frequencies from perturbations of equatorial circular 
geodesics. Semer\'{a}k and \v{Z}\'{a}\v{c}ek \cite{Semerak} have done such calculations for the 
superposition of a Schwarzschild black hole with the Lemos-Letelier disk. They found 
that heavier disks are more stable with respect to horizontal perturbations near their 
inner rims, whereas they are less stable with respect to vertical perturbations. For 
astrophysical relevance, it is important to determine not only the stability of circular 
motion on the disk plane, but also stability in the vertical direction.

Using the perturbed equations for equatorial circular geodesics, the epicyclic frequency with respect to 
infinity $\omega_h$ and the vertical oscillation frequency with respect to 
infinity $\omega_v$ for the metric (\ref{eq_weyl_metric}) are given by 
(see \cite{Semerak} for a detailed deduction)
\begin{align}
\omega_h^2 &=\frac{e^{2\phi-\nu}}{2-r\phi_{,r}}\left( \phi_{,rr}+r\phi_{,r}^3-
3\phi_{,r}^2+\frac{3}{r}\phi_{,r} \right) \mbox{,} \label{eq_osc_h} \\
\omega_v^2 &=\frac{e^{2\phi-\nu}}{2-r\phi_{,r}}\left[ \phi_{,zz}-2\phi_z^2
(1-r\phi_{,r}) \right] \mbox{.} \label{eq_osc_v}
\end{align}
In Eq.\ (\ref{eq_osc_v}) the function $\phi_z$ is obtained from the 
limit $\lim_{z \rightarrow 0^{\pm}} \phi_z$ and $\phi_{,zz}$ follows 
from Eq.\ (\ref{eq_eins1}). Stable horizontal and vertical orbits are 
only possible where $\omega_h^2>0$ and $\omega_v^2>0$, 
respectively. Note that condition $\omega_h^2>0$ is equivalent to condition (\ref{eq_stab}) 
which follows from Rayleigh stability criteria. 

\begin{figure}
\centering
\includegraphics[scale=0.8]{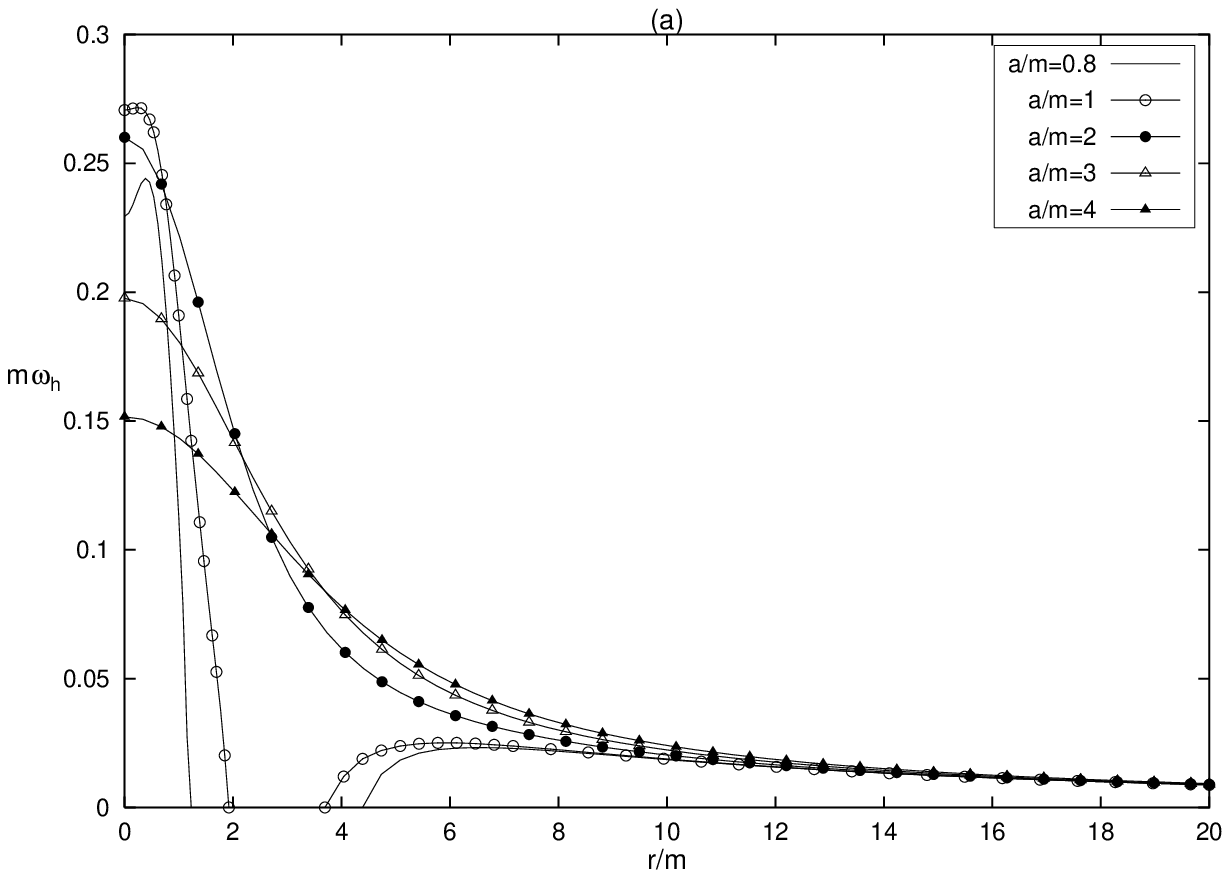}\\
\includegraphics[scale=0.8]{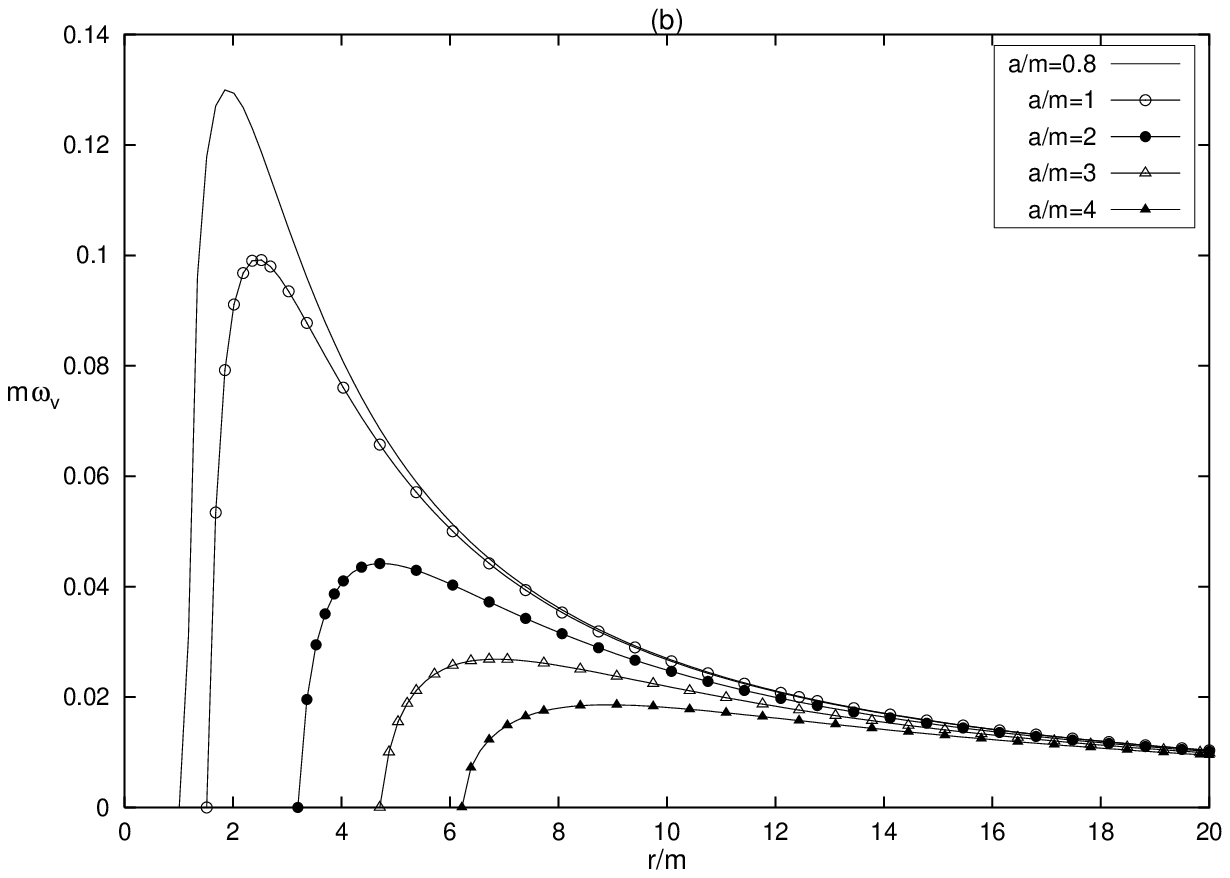}
\caption{Radial profiles of (a) the epicyclic frequency $m\omega_h$ and (b) the vertical 
oscillation frequency $m\omega_v$ for an isolated Chazy-Curzon disk.} \label{fig_5}
\end{figure}

We compute first the frequencies for an isolated Chazy-Curzon disk, since is seems that 
such a calculation has not been done before for this class of disks. In Fig.\ \ref{fig_5}(a) we 
plot the epicyclic frequency as functions of radius $r/m$ and cut parameter $a/m$. For 
$a/m > 1.015$ the disks always are stable and the epicyclic frequency is lowered for less relativistic 
disks. Highly relativistic disks (curve with $a/m=0.8$ for example) develop annular regions of instability. 
The curves of vertical oscillation frequencies are depicted in Fig.\ \ref{fig_5}(b). We note that in this case 
highly relativistic disks are more stable in the vertical direction. In Eq.\ (\ref{eq_osc_v}) the term with 
$\phi_z$ is small compared to $\phi_{,zz}$, thus if we consider only 
\begin{equation}
\phi_{,zz}=\frac{2m}{(r^2+a^2)^{5/2}}(r^2-2a^2) \mbox{,}
\end{equation}
we note that vertical oscillations are zero at $r=a\sqrt{2}$, so the regions of vertical stability 
are enlarged as the cut parameter $a$ is decreased.

\begin{figure}
\centering
\includegraphics[scale=0.8]{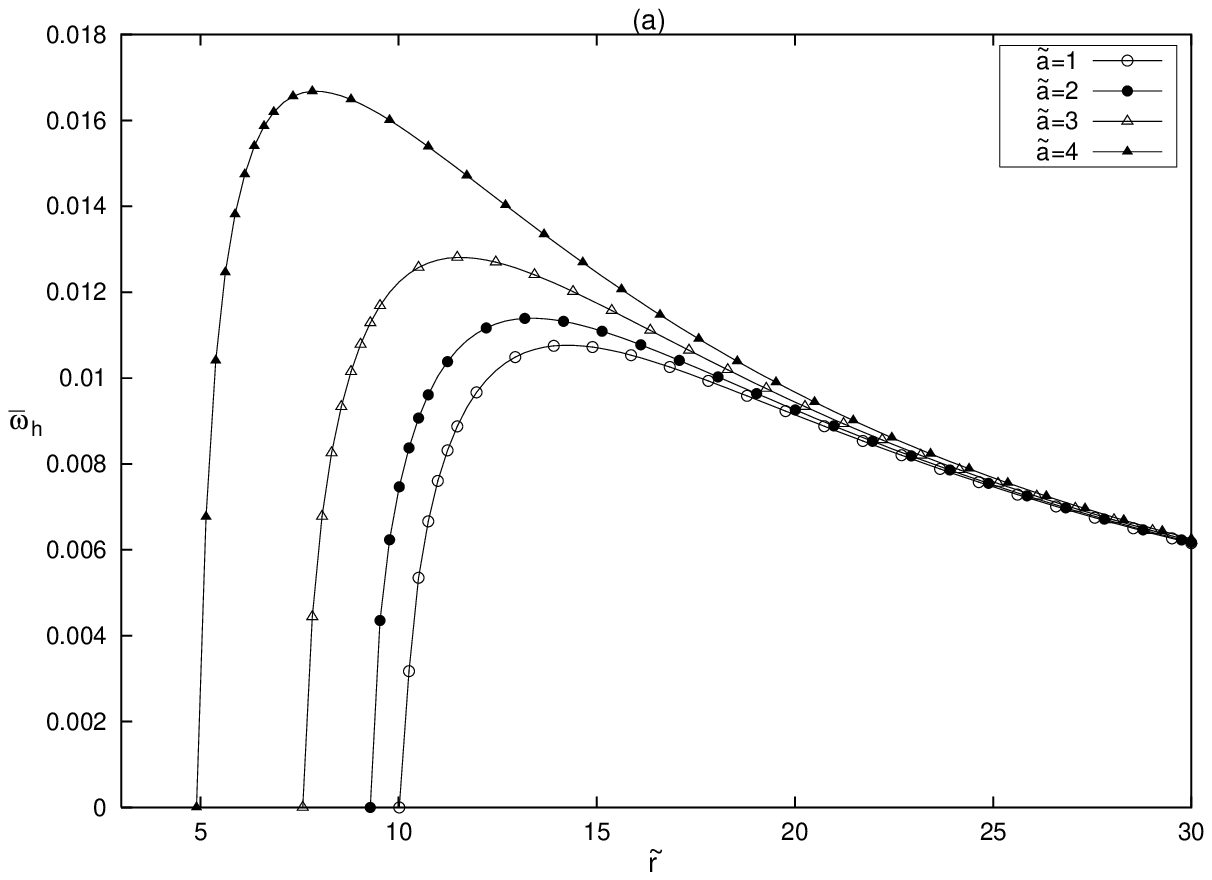}\\
\vspace{0.2cm}
\includegraphics[scale=0.8]{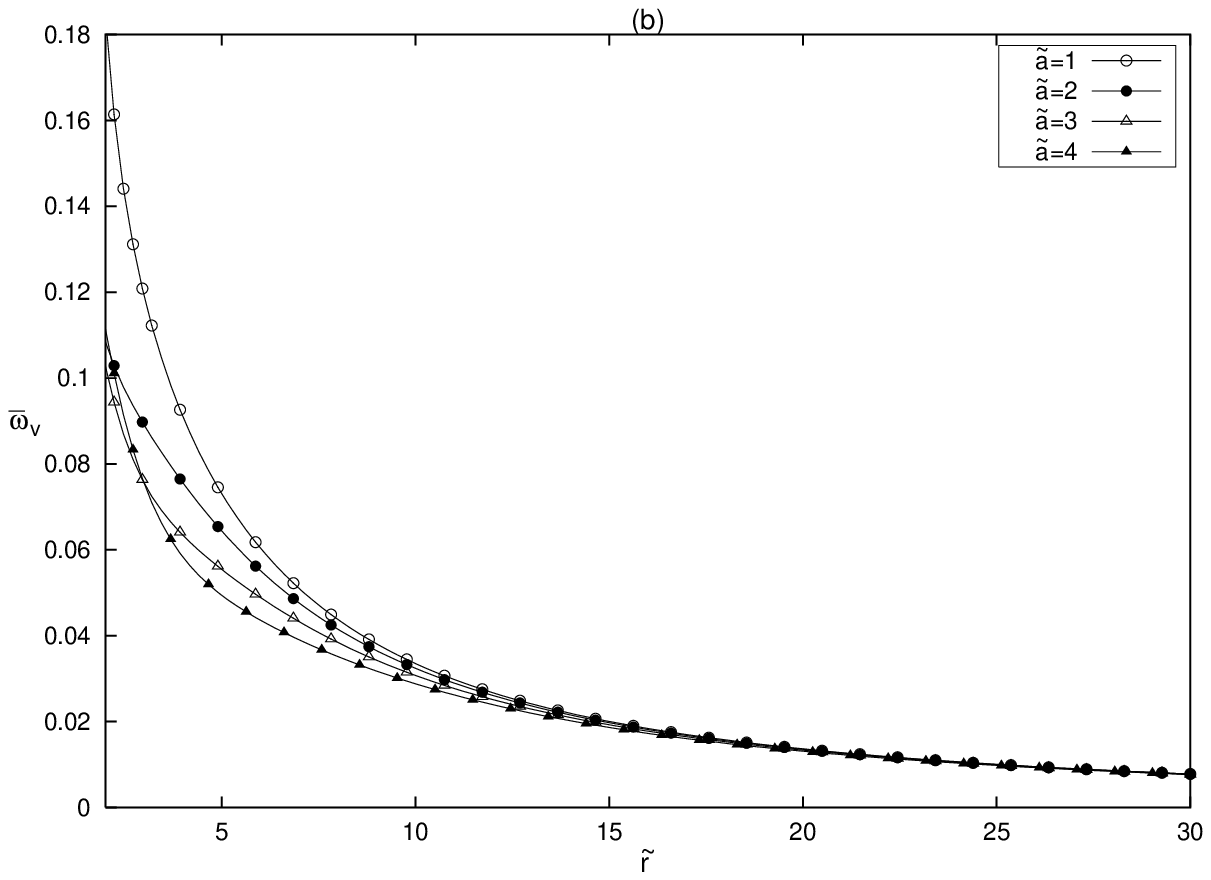}
\caption{Radial profiles of (a) horizontal and (b) vertical oscillation frequencies
of Chazy-Curzon disk with a black hole without rods. Parameters: $\alpha=1$, $\tilde{a}=1$, 
$2$, $3$ and $4$.} \label{fig_6}
\end{figure}

Now we consider the superposition of a Curzon disk and a black hole \emph{without} rods.
Fig.\ \ref{fig_6} shows curves of (a) horizontal $\bar{\omega}_h=M\omega_h$ and (b) vertical 
$\bar{\omega}_v=M\omega_v$ oscillation
frequencies of the disk with $\alpha=1$ and four different values of the ``cut'' parameter 
$\tilde{a}$. Now we always have regions of horizontal instability that begin at the innermost 
stable circular orbit and decrease for less relativistic disks. With respect to vertical oscillations,
 it is seen from Fig.\ \ref{fig_6}(b) that there are no regions of vertical instabilities. Thus one can 
conclude that the black hole desestabilizes the Curzon disk in the horizontal direction, whereas the 
opposite is true for the vertical direction. 

\begin{figure}
\centering
\includegraphics[scale=0.8]{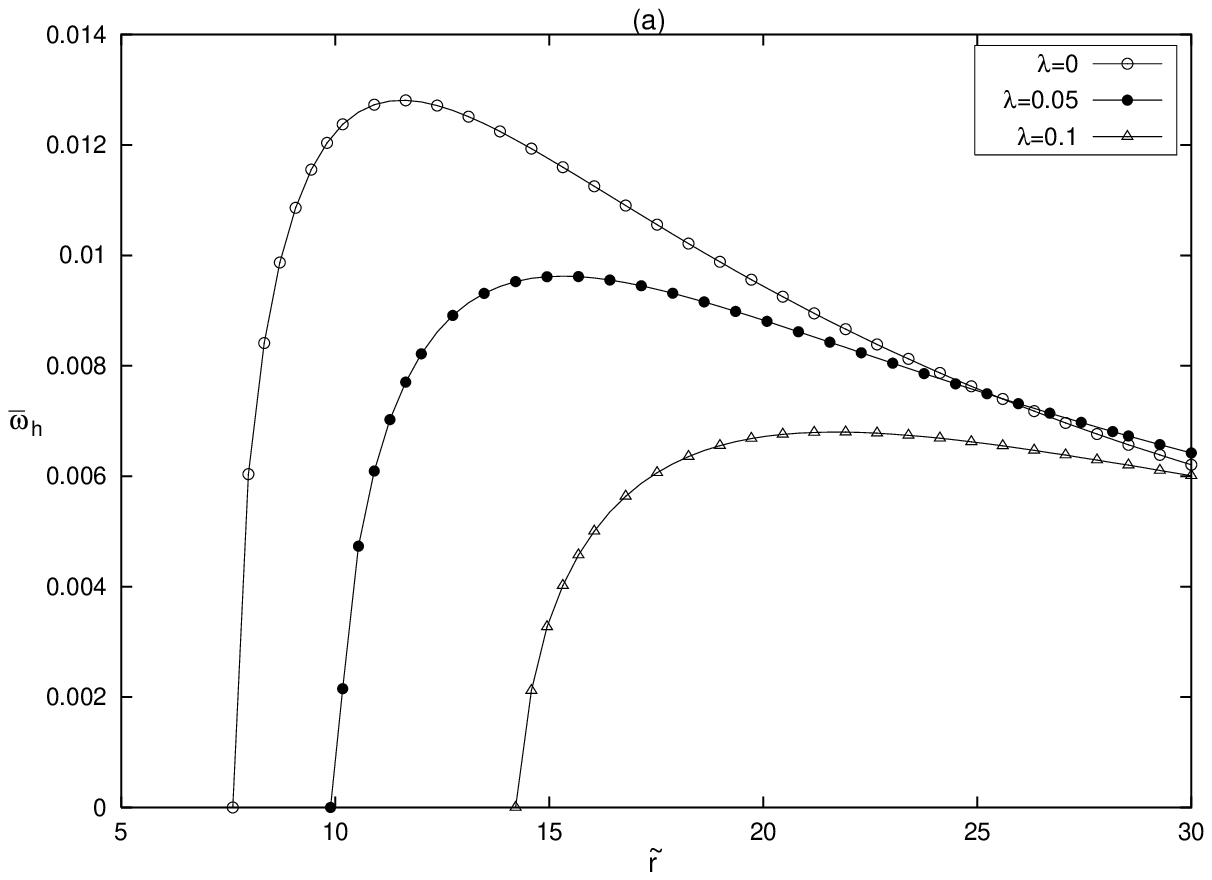}\\
\vspace{0.2cm}
\includegraphics[scale=0.8]{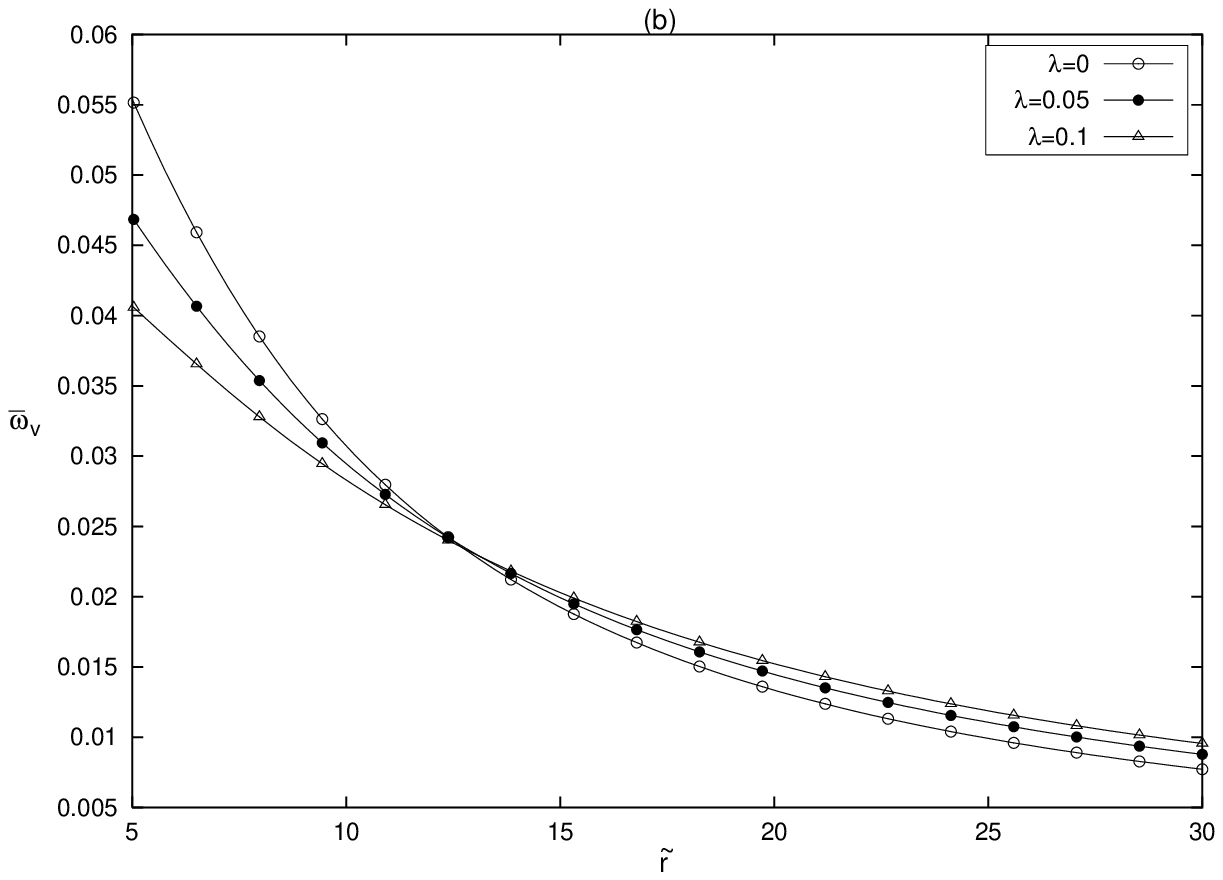}
\caption{Radial profiles of (a) horizontal and (b) vertical oscillation frequencies
of Chazy-Curzon disk with black hole and rods. Parameters are $\tilde{c}_2=11$,
$\alpha=1$, $\tilde{a}=3$, $\lambda=0$, $0.05$ and $0.1$.} \label{fig_7}
\end{figure}
In Fig.\ \ref{fig_7}(a)-(b) we graph again $\bar{\omega}_h=M\omega_h$ and $\bar{\omega}_v=M\omega_v$, respectively, for the Chazy-Curzon disk with black hole and rods, 
with their length fixed and vary the linear mass density $\lambda$. As expected from the curves of
Fig.\ \ref{fig_3}(b), the more massive the rods, the larger are the disk's unstable regions in the 
horizontal direction. The rods also tend to lower the vertical oscillation frequencies near the 
disk's center, but unstable regions do not appear.
\section{Geodesic orbits} \label{sec_orbits}

\begin{figure}
\centering
\includegraphics[scale=0.8]{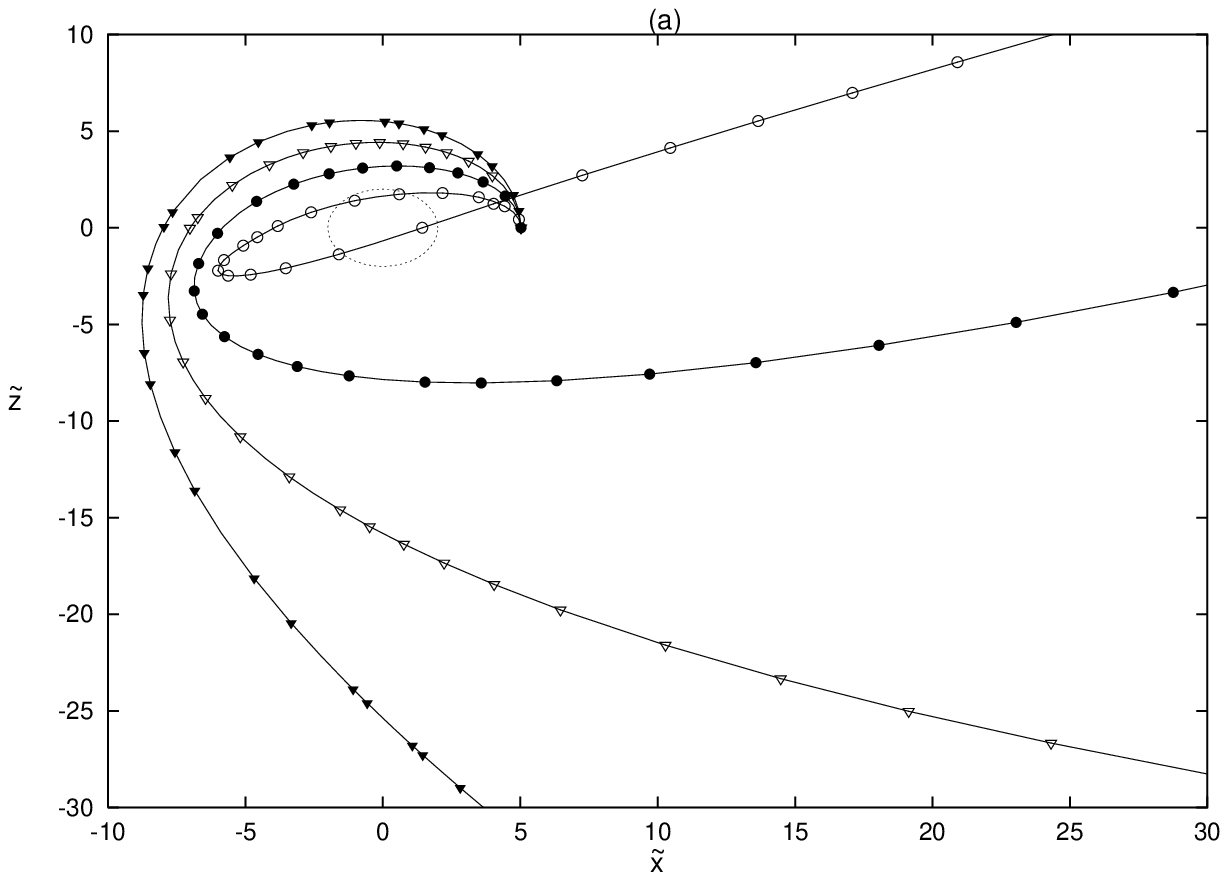}\\
\includegraphics[scale=0.8]{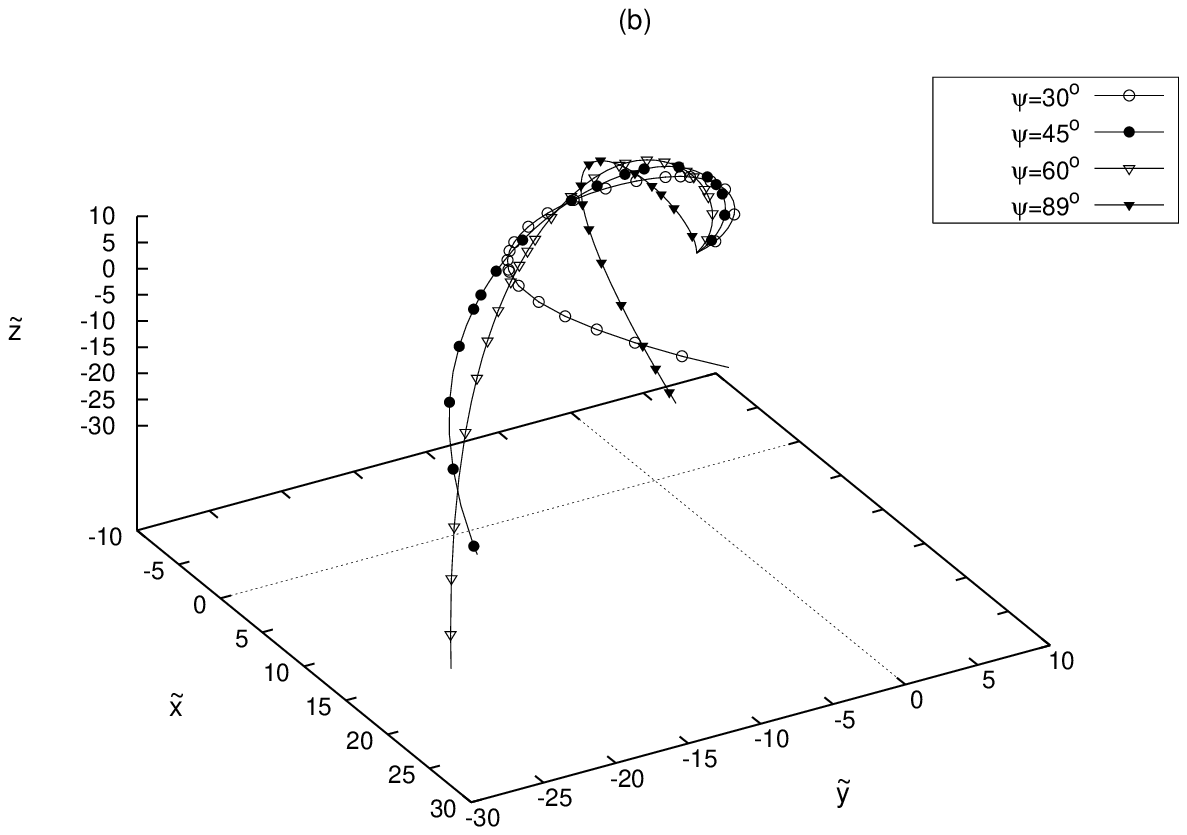}
\caption{Geodesic orbits for the superposition Curzon disk + black hole without rods. 
Parameters: $\alpha=1$, $\tilde{a}=3$, $\mathcal{E} \approx 1.01$, $\tilde{r}_0=3.9$. (a) Projection 
on the $x-z$ plane of the curves in (b).} \label{fig_8}
\end{figure}

In the previous section perturbations of equatorial circular geodesics
were used to discuss disk stability for the system disk + black hole + rods. 
Now we solve numerically the geodesic equations of motion 
\begin{equation}
\ddot{x}^{\mu}+\Gamma^{\mu}_{\alpha \beta}\dot{x}^{\alpha}\dot{x}^{\beta}=0 \mbox{,}
\end{equation}
for metric Eq.\ (\ref{eq_weyl_metric}), where $\Gamma^{\mu}_{\alpha \beta}$ are
the Christoffel symbols and the dot denote differentiation with respect to the 
proper time. Defining the orthonormal tetrad 
$\{ V^a,W^a,Y^a,Z^a \}$ where
\begin{subequations}
\begin{align}
V^a &=e^{-\phi/2}(1,0,0,0) \mbox{,}\\
W^a &=e^{(\phi-\nu)/2}(0,1,0,0) \mbox{,}\\
Y^a &=e^{(\phi-\nu)/2}(0,0,1,0) \mbox{,}\\
Z^a &=\frac{e^{\phi/2}}{r}(0,0,0,1) \mbox{,}
\end{align}
\end{subequations}
the tetrad components of the four-velocity $v^a$ read 
\begin{equation}
v^a=\gamma(1,v\sin \psi \cos \chi,v\sin \psi \sin \chi,v\cos \psi) \mbox{,}
\end{equation}
with $\gamma=1/\sqrt{1-v^2}$. The specific energy and angular momentum of the test 
particle are
\begin{align}
\mathcal{E} &=e^{\phi}\dot{t}=e^{\phi/2}\gamma \mbox{,} \\
h &=r^2e^{-\phi}\dot{\varphi}=re^{-\phi/2}\gamma v \cos \psi \mbox{.}
\end{align} 
As initial conditions we take a position at radius $r_0$ on the disk's plane and components of the 
four-velocity $v_0^a=\gamma(1,0 ,v_0\sin \psi,v_0\cos \psi)$, where $v_0$ is equal to 
the tangential velocity of circular orbits at radius $r_0$. We choose initial radii such that the 
energy is slightly higher than the escape energy. Fig.\ \ref{fig_8}(a)-(b) shows the orbits of particles 
in the presence of the black hole and Curzon disk without rods. The parameters are $\alpha=1$, 
$\tilde{a}=3$, $\tilde{r}_0=3.9$, $\mathcal{E} \approx 1.01$ and different initial angles $\psi$. 
Fig.\ \ref{fig_8}(a) is a projection of the orbits on the $x-z$ plane. The coordinates have been 
transformed from Weyl to Schwarzschild coordinates $(t,\mathsf{r},\theta,\varphi)$ via the relations
\begin{equation}
r=\sqrt{\mathsf{r}(\mathsf{r}-2M)} \sin \theta, \qquad z=(\mathsf{r}-M) \cos \theta \mbox{,}
\end{equation}
and then to $x=\mathsf{r} \sin \theta \cos \varphi$, $y= \mathsf{r} \sin \theta \sin \varphi$ and 
$z=\mathsf{r} \cos \theta$. 

\begin{figure}
\centering
\includegraphics[scale=0.6]{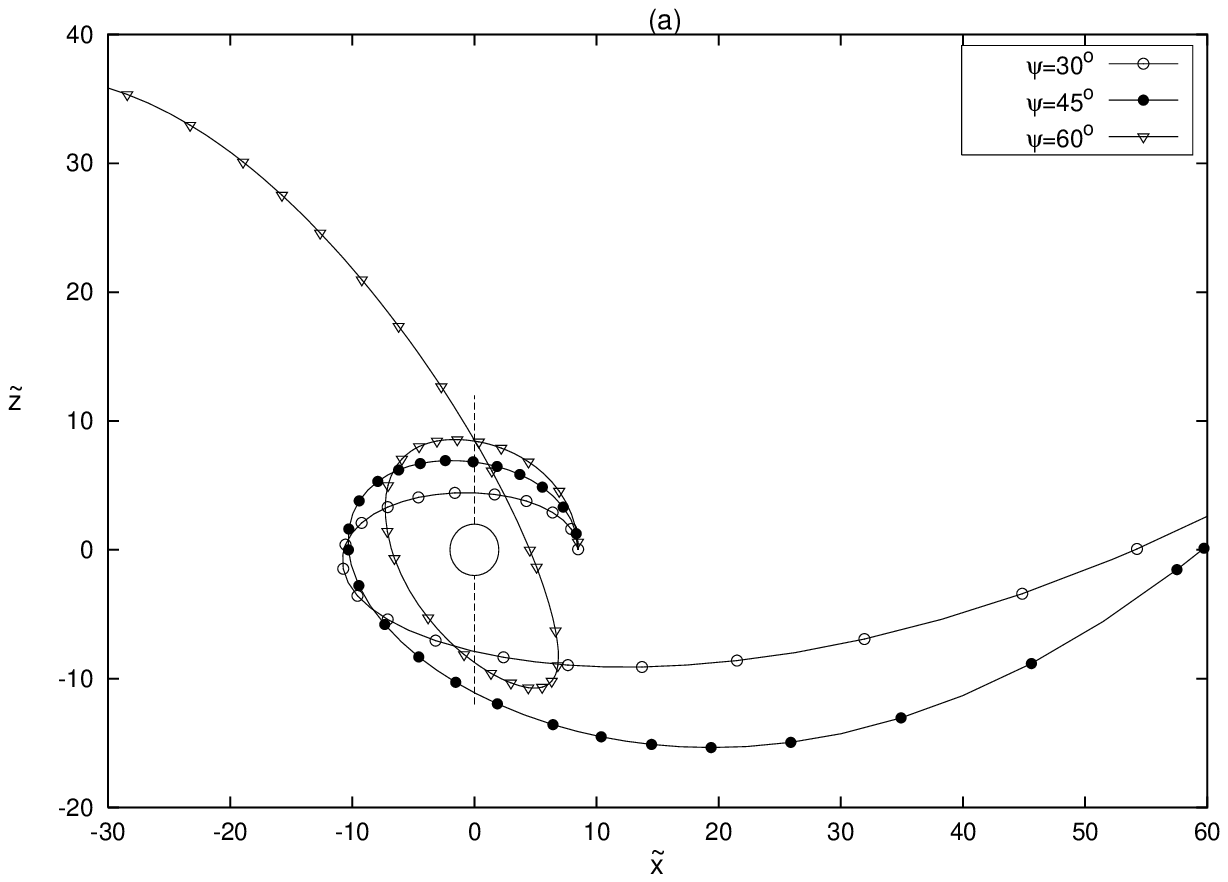}%
\includegraphics[scale=0.6]{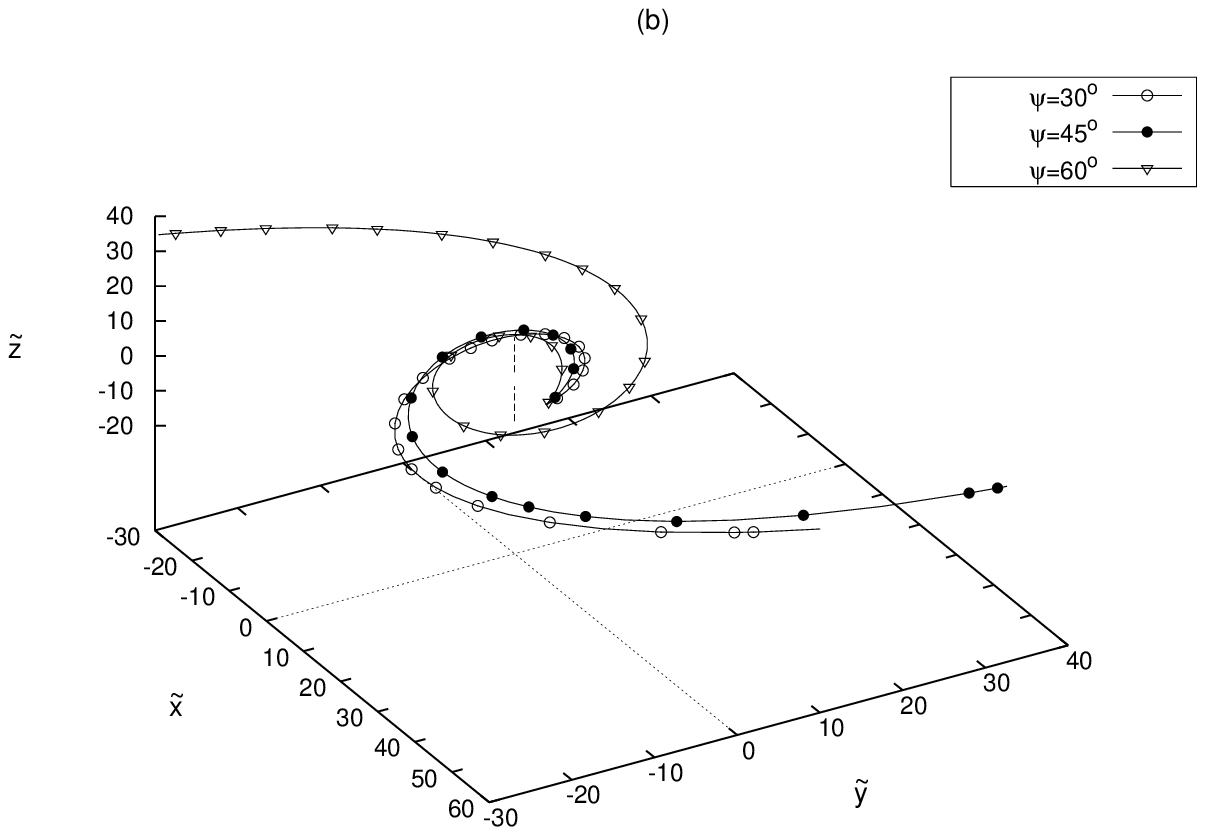}\\
\includegraphics[scale=0.6]{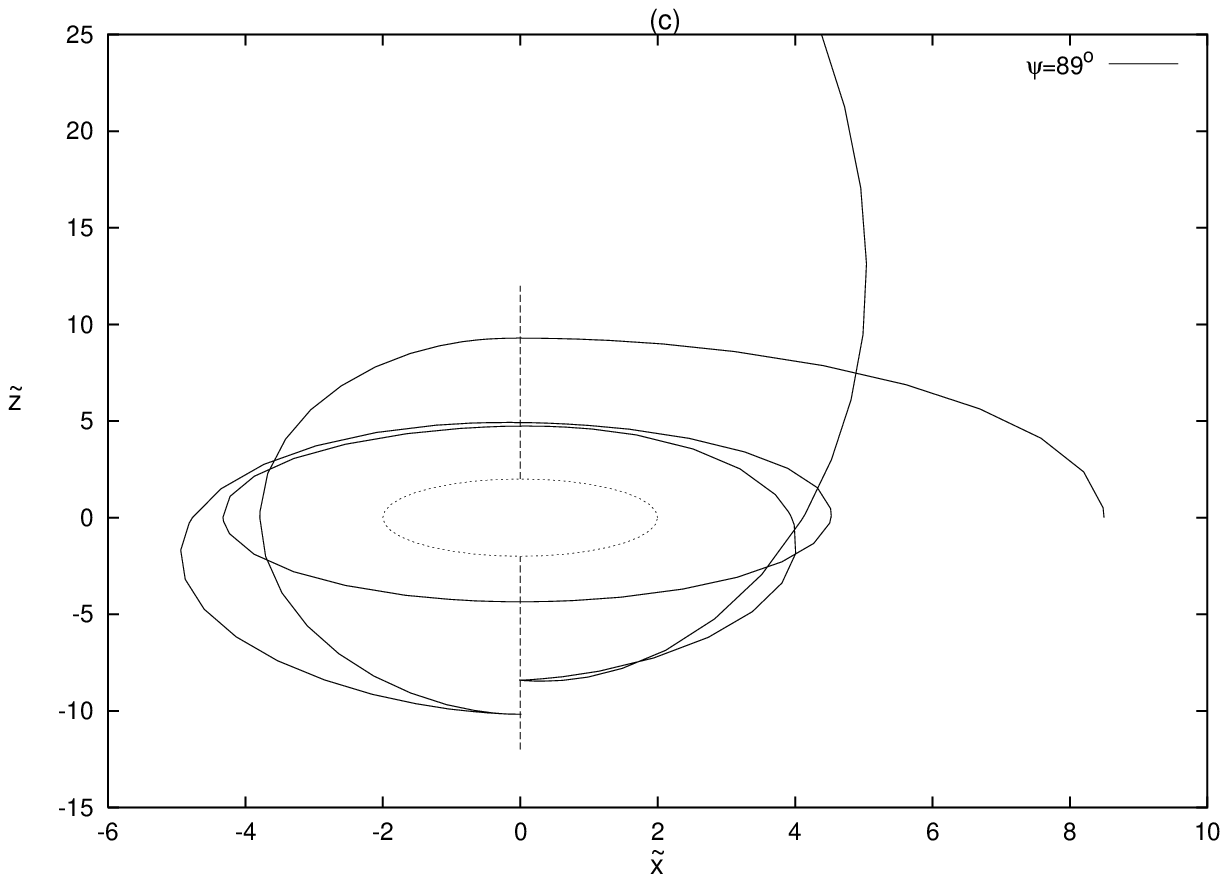}%
\includegraphics[scale=0.6]{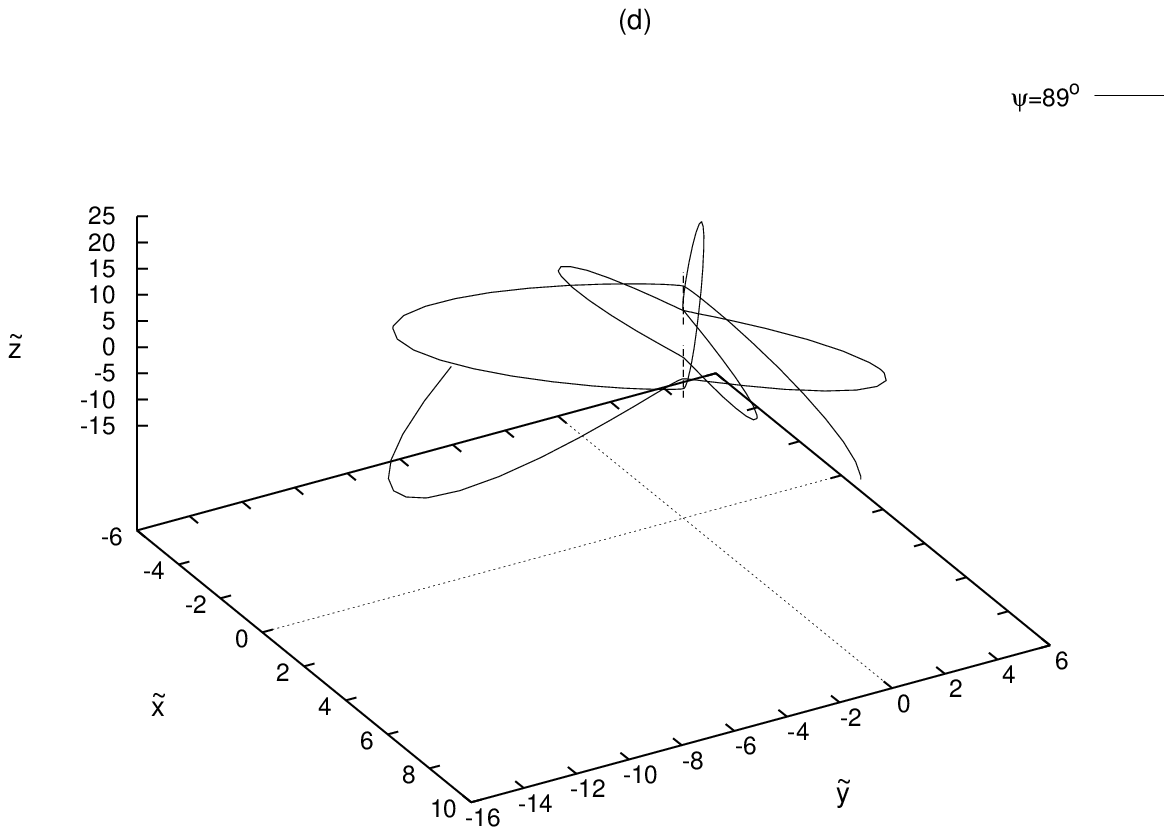}
\caption{ Geodesic orbits for the superposition Curzon disk + black hole with rods. 
Parameters: $\alpha=1$, $\tilde{a}=3$, $\lambda=0.1$, $\tilde{c}_2=11$, $\mathcal{E} \approx 1.01$ 
$\tilde{r}_0=7.43$. The curve for $\psi=89^o$ is displayed in (c) and (d) for better visualization.} \label{fig_9}
\end{figure}

In Fig.\ \ref{fig_9}(a)-(d) we have computed some orbits now with the rods. The parameters are 
$\alpha=1$, $\tilde{a}=3$, $\tilde{r}_0=7.43$, $\lambda=0.1$, $\tilde{c}_2=11$, $\mathcal{E} \approx 1.01$ and 
different initial angles $\psi$. The orbit with $\psi=89^o$ has been placed in a separate graph for 
better visualization. For low initial angles, the rods have little effect on the trajectories, but this is not 
true as the particles approach the $z$ axis. The orbit in Fig.\ \ref{fig_9}(c)-(d) even suggests that we can 
expect chaotic behaviour for orbits that pass very near the rods. In fact, it has been 
shown \cite{Gueron} that prolate quadrupole deformations can introduce chaotic motion of geodesic
test particles. In the oblate case, only regular motion was found.

\section{Discussion} \label{sec_discuss}

We presented a very simplified, although exact, general relativistic model of an 
active galactic nuclei based on a 
superposition of a Schwarzschild black hole, a Chazy-Curzon disk and two rods 
placed on the symmetry axis, representing jets. We found that the presence of the 
rods enhances the disk regions with superlumial velocities. Using an extension of 
Rayleigh criteria of stability, it was found that in general the rods also increase the
regions of instability, but when the rods are large and the disk's mass is low they can contribute
to stabilize the disk. Also disk stability in the vertical direction was studied through perturbation
of circular geodesics. The rods contribute to lower the vertical oscillation frequencies near
the disk's center. Some geodesic orbits calculated numerically for the system
black hole + disk + rods show the possibility of chaotic trajectories near the rods.

The model here presented should be viewed as a first approach. As was stated in the introduction, more realistic models of active galactic nuclei 
should incorporate rotation and electromagnetic fields. However, the analysis of such a model
would not be trivial, because of the large number of free parameters involved.

\bigskip
\centerline{{\large \textbf{Acknowledgments}}}

\bigskip
D.\ V.\ thanks CAPES for financial support. P.\ S.\ L.\ thanks CNPq and FAPESP 
for financial support. 
\section{Appendix}

The metric function Eq.\ (\ref{eq_eins2}) for the superposition of a black hole, two 
rods and a disk generated from Chazy-Curzon solution can be calculated as follows. We 
rewrite potential Eq.\ (\ref{eq_phi_sup}) as 
\begin{equation}
\phi=\phi_{R1}+\phi_{BH}+\phi_{R2}+\phi_{D}^{+} \mbox{,}
\end{equation}
where
\begin{align*}
\phi_{R1} &= -2\lambda \ln \left( \frac{\mu_3}{\mu_4} \right) \mbox{,}
& \phi_{BH} = -2\lambda \ln \left( \frac{\mu_1}{\mu_2} \right) \\
\phi_{R2} &= -2\lambda \ln \left( \frac{\mu_5}{\mu_6} \right) \mbox{,}
& \phi_{D}^{+} =\frac{m}{\epsilon} \ln \left( \frac{\mu_7}{\mu_8} \right) \mbox{,}
\end{align*}
with $\mu_7=-a-\epsilon-z+\sqrt{r^2+(-a-\epsilon-z)^2}$ and $\mu_8=-a+\epsilon-z+
\sqrt{r^2+(-a+\epsilon-z)^2}$. In the limit $\epsilon \rightarrow 0$ expression for 
$\phi_{D}^{+}$ reduces to the Chazy-Curzon disk  Eq.\ (\ref{eq_curzon}) on $z>0$. Thus 
all terms can be expressed as $\mu$ potentials. Using repeatedly properties 
(\ref{eq_nu_rel2})-(\ref{eq_ident}) we get
\begin{align}
\nu & [\phi_{R1}+\phi_{BH}+\phi_{R2}+\phi_{D}^{+}]=\nu [\phi_{R1}]+\nu [\phi_{BH}]+\nu [\phi_{R2}]+\nu [\phi_{D}^+] \notag \\
& +2\nu [\phi_{R1},\phi_{BH}]+2\nu [\phi_{R1},\phi_{R2}]+
2\nu [\phi_{R1},\phi_{D}^+]+2\nu [\phi_{BH},\phi_{R2}] \notag \\
& +2\nu [\phi_{BH},\phi_{D}^+]+2\nu [\phi_{R2},\phi_{D}^+] \mbox{,} \label{eq_nu_sup}
\end{align}
with
\begin{align*}
\nu &[\phi_{R1}]= 4\lambda^2 \ln \left[ \frac{(r^2+\mu_3 \mu_4)^2}{(r^2+\mu_3^2)(r^2+\mu_4^2)} 
\right] \mbox{,} \\
\nu &[\phi_{BH}]= \ln \left[ \frac{(r^2+\mu_1 \mu_2)^2}{(r^2+\mu_1^2)(r^2+\mu_2^2)} 
\right] \mbox{,} \\
\nu &[\phi_{R2}]= 4\lambda^2 \ln \left[ \frac{(r^2+\mu_5 \mu_6)^2}{(r^2+\mu_5^2)(r^2+\mu_6^2)} 
\right] \mbox{,} \\
\nu & [\phi_{D}^{+}]=-\frac{m^2r^2}{[r^2+(z+a)^2]^2} \mbox{,} \\
\nu &[\phi_{R1},\phi_{BH}]=2\lambda \ln \left[ \frac{(r^2+\mu_1\mu_3)(r^2+\mu_2\mu_4)}{(r^2+\mu_1\mu_4)(r^2+\mu_2\mu_3)} \right] \mbox{,} \\
\nu &[\phi_{R1},\phi_{R2}]=4\lambda^2 \ln \left[
\frac{(r^2+\mu_3\mu_6)(r^2+\mu_4\mu_5)}{(r^2+\mu_3\mu_5)(r^2+\mu_4\mu_6)} \right] \mbox{,} \\
\nu &[\phi_{R1},\phi_{D}^+]=\frac{2\lambda m}{(a+c_1)(a+c_2)\sqrt{r^2+(a+z)^2}} \left[ (a+c_2)\sqrt{r^2+(c_1-z)^2} \right. \notag \\
& \left. -(a+c_1)\sqrt{r^2+(c_2-z)^2}+(c_1-c_2)\sqrt{r^2+(a+z)^2} \right] \mbox{,}\\
\nu &[\phi_{BH},\phi_{R2}]=2\lambda \ln \left[ \frac{(r^2+\mu_1\mu_5)(r^2+\mu_2\mu_6)}{(r^2+\mu_1\mu_6)(r^2+\mu_2\mu_5)} \right] \mbox{,} \\
\nu &[\phi_{BH},\phi_{D}^+]=\frac{m}{(a^2-M^2)\sqrt{r^2+(a+z)^2}} \left[ (a+M)\sqrt{r^2+(M+z)^2} \right. \notag \\
& \left. -(a-M)\sqrt{r^2+(M-z)^2}-2M\sqrt{r^2+(a+z)^2}\right] \mbox{,}\\
\nu &[\phi_{R2},\phi_{D}^+] =\frac{2\lambda m}{(a-c_1)(a-c_2)\sqrt{r^2+(a+z)^2}} \left[ (c_2-a)\sqrt{r^2+(c_1+z)^2} \right. \notag \\
& \left. -(c_1-a)\sqrt{r^2+(c_2+z)^2}+(c_1-c_2)\sqrt{r^2+(a+z)^2} \right]  \mbox{.}
\end{align*}
In the particular case $c_1=M$ $(\mu_4=\mu_2$ and $\mu_5=\mu_1)$, and on $z=0$, Eq.\ (\ref{eq_nu_sup}) simplifies to
\begin{align}
\nu &[\phi_{R1}+\phi_{BH}+\phi_{R2}+\phi_{D}^{+}]= \notag \\
& \ln \left[
\frac{r^{16 \lambda^2-8\lambda+2}(r^2+\mu_1\mu_6)^{8\lambda^2-4\lambda}(r^2+\mu_2\mu_3)^{8\lambda^2-4\lambda}}
{(r^2+c_2^2)^{4\lambda^2}(r^2+M^2)^{4\lambda^2-4\lambda+1}(r^2+\mu_1\mu_3)^{8\lambda^2-4\lambda}(r^2+\mu_2\mu_6)^{8\lambda^2-4\lambda}} \right] \notag \\
& -\frac{m^2r^2}{(r^2+a^2)^2}+\frac{8\lambda m}{(a^2-c_2^2)(a^2-M^2)\sqrt{r^2+a^2}} \left[ c_2(a^2-M^2)\sqrt{r^2+c_2^2} \right. \notag \\
& \left. -M(a^2-c_2^2)\sqrt{r^2+M^2}+(M-c_2)(a^2+Mc_2)\sqrt{r^2+a^2} \right] \notag \\
&+4mM\frac{(\sqrt{r^2+M^2}-\sqrt{r^2+a^2})}{(a^2-M^2)\sqrt{r^2+a^2}} \mbox{.}
\end{align}

\end{document}